\documentstyle[12pt]{article}
\begin{document}
\title{{\bf SENSIBLE QUANTUM MECHANICS:
\\ARE ONLY PERCEPTIONS PROBABILISTIC?}
\thanks{Alberta-Thy-05-95, quant-ph/9506010}}
\author{
Don N. Page
%\thanks{CIAR  Cosmology Fellow}
\thanks{Internet address:
don@phys.ualberta.ca}
\\
CIAR Cosmology Program, Institute for Theoretical Physics\\
Department of Physics, University of Alberta\\
Edmonton, Alberta, Canada T6G 2J1
}
\date{(1995 June 7, revised 1997 June 30)}

\maketitle
\large
\begin{abstract}
\baselineskip 14 pt

 Quantum mechanics may be formulated as
{\it Sensible Quantum Mechanics} (SQM)
so that it contains nothing probabilistic,
except, in a certain frequency sense, conscious perceptions.
Sets of these perceptions can be deterministically realized
with measures given by expectation values of
positive-operator-valued {\it awareness operators}
in a quantum state of the universe which never jumps or collapses.
Ratios of the measures for these sets of perceptions
can be interpreted as frequency-type probabilities
for many actually existing sets rather than as propensities
for potentialities to be actualized,
so there is nothing indeterministic in SQM.
These frequency-type probabilities generally cannot be given
by the ordinary quantum ``probabilities'' for a single set of alternatives.
{\it Probabilism}, or ascribing probabilities to  unconscious aspects
of the world, may be seen to be an {\it aesthemamorphic myth}.

 No fundamental correlation or equivalence is postulated between different
perceptions (each being the entirety of a single conscious experience and thus
not in direct contact with any other), so SQM, a variant of Everett's
``many-worlds'' framework, is a ``many-perceptions'' framework but not a
``many-minds'' framework.  Different detailed SQM theories may be tested
against experienced perceptions by the {\it typicalities} (defined herein) they
predict for these perceptions.  One may adopt the {\it Conditional Aesthemic
Principle}:  among the set of all conscious perceptions, our perceptions are
likely to be typical.

 An experimental test is proposed to compare SQM with a variant, SQM$n$.
\\
\\
\end{abstract}
\normalsize
\baselineskip 15 pt

\section{Basic Ideas of Sensible Quantum Mechanics}

\hspace{.25in}Probabilities can seem rather mysterious in any theory or
description that is supposed to be complete.  If probabilities are interpreted
as indicating fundamental uncertainties, then any theory that describes things
probabilistically appears to be uncertain and incapable of being as complete as
some alternative theory that describes more precisely what happens.

 In many cases the incompleteness of a theory is overbalanced in our evaluation
by the relative simplicity it has in comparison with a more complete theory.
Thus in classical physics, for example, we may consider a statistical theory
better than a more complete alternative which gives the precise trajectory in
phase space, because the statistical theory may be much simpler.  However, in
such cases we readily agree that the statistical theory is incomplete and
usually believe that a more complete description exists in principle.

 Someone may despair of ever knowing a complete description of a certain system
(e.g., the complete history of the universe), and it may indeed be true that he
will never succeed in finding it, but that personal despair should not be
misinterpreted as evidence that no such complete description exists in
principle.  Furthermore, one can argue that a complete description certainly
exists, namely, the system itself.  (One might prefer that a complete theory
appear simpler than the system it describes, but that is a separate question
from that of the {\it existence} of a complete theory.)  So in this paper I
shall assume that a complete theory of the universe does exist in principle,
and that it is a goal of physics to search for one or at least to try to get
closer to one.

 In the absence of a better alternative, it seems worthwhile to consider
whether quantum mechanics is a suitable framework for a complete theory of the
universe.  But then one runs into the problem that quantum mechanics is usually
interpreted probabilistically, which seems to indicate that it cannot be
complete.  For example, if the quantum probabilities are interpreted as
propensities for several possible sequences of events to be actualized, and if
only one sequence actually occurs, then the theory is incomplete in not
describing which particular sequence does occur.

 Here, expanding upon previous work \cite{Page}, I argue that
quantum mechanics {\it is} a complete framework for describing the unconscious
aspects of reality (here called the {\it quantum world}), because those aspects
should {\it not} be described probabilistically.  Instead, I claim it is
consistent to assume that they are completely described by the quantum
amplitudes (e.g., by the amplitudes in a path integral in a sum-over-histories
approach, or by an algebra of operators and a quantum state giving the
expectation value of these operators).

 (More modestly, I am really merely claiming that quantum mechanics {\it may
be} a complete framework {\it if} it is absolutely correct.  In view of the
progression of science, that latter assumption may be thought to be highly
dubious, but in the spirit of what Feynman \cite{Fey} called Wheeler's
``radical conservatism,'' I want to push as far as possible our present
principles, in this case the assumption that quantum mechanics {\it is}
correct, which I shall make throughout the remainder of this paper.  I wish to
transcend Wald's insightful remark on page 309 of \cite{Pen}, ``If you believe
in quantum mechanics, then you can't take it seriously.''  In particular, I am
arguing that the objections that can be raised against a probabilistic theory
need not be objections against quantum mechanics in the version I am
proposing.)

 However, our universe also includes conscious perceptions or sensations
or phenomenal first-person experiences (in what is here
called the {\it conscious world} to distinguish it from the unconscious {\it
quantum world}, though both together can be taken to be the {\it physical
world} if one accepts that terminology; this {\it physical world} includes both
the ``mental world'' and the ``physical world'' of Fig. 8.1 of Penrose's {\em
Shadows of the Mind} \cite{Pen}).  These perceptions seem to have certain
classical aspects that are not captured merely by the full set of amplitudes
for the quantum world, so I propose that quantum mechanics be augmented to give
real measures for sets of conscious perceptions.  In the particular
augmentation I am proposing, which I call {\it Sensible Quantum Mechanics}, or
SQM, these measures are given by the {\it expectation values} of particular
{\it awareness operators} in the state of the quantum world.  (This strong
connection between the conscious world and the quantum world in a thus-unified
physical world means that Sensible Quantum Mechanics is not fundamentally
dualistic in any negative sense, any more than the quantum world is dualistic
for having the distinct elements of paths and amplitudes, or of operators and a
quantum state.  However, it is dualistic in the sense of Chalmer's
{\em The Conscious Mind} \cite{Chal},
with which I agree virtually completely for the first half of the book,
except for some minor issues of terminology, and for the final chapter, on
the interpretation of quantum mechanics,
even though I had not seen this excellent book
when I wrote the present paper.)

 Ratios of the measures of appropriate sets of conscious perceptions can be
interpreted as the classical conditional probabilities for these sets.  In this
way Sensible Quantum Mechanics gives something like the usual probabilistic
interpretation applied to ordinary quantum mechanics, but I am proposing that
in the most fundamental sense, probabilities are entirely restricted to
conscious perceptions or sensations.  One might summarize this proposal by the
slogan, ``No nonsensical probabilities!''

 Thus I am proposing a framework or viewpoint in which {\it probabilism}, or
interpreting the unconscious quantum world itself probabilistically, is an {\it
aesthemamorphism} (from the Greek
$\alpha\iota\sigma\theta\eta\sigma\iota\sigma$:  perception, sense, sensation).
 It is a myth of attributing a fundamental property of conscious perceptions to
the quantum world, rather analogous to the myth of animism that ascribes living
properties to inanimate objects.  Of course, probabilism may be a convenient
myth, just as animism is a convenient myth when we say such things as, ``A
charged particle {\it feels} an electromagnetic field,'' or when Feynman \cite{FeyLec}
wrote, ``It isn't that a particle takes the path of least action
but that it smells all the paths in the neighborhood \ldots,'' but it would give us a
better understanding of the world if we recognized it as a myth.  My claim that
probabilism is a myth is also analogous to the claim that classical physics is
a myth, since it is only an approximation to an underlying quantum reality.
(Of course, it may be that quantum mechanics itself is a myth, but, as
discussed above, in this paper I am making the radically conservative
assumption that it is absolutely correct.  My claim is that I have good reasons
for identifying probabilism as a myth, whereas I do not see any good evidence
yet that quantum mechanics itself is a myth.)

 Thus, by eliminating probabilities from the quantum world, quantum mechanics
is permitted to be a complete framework for that world.  On the other hand,
someone may object that Sensible Quantum Mechanics reintroduces probabilities
for the conscious world and so cannot possibly be a complete theory for the
entire physical world of both the quantum world and the conscious world.

 This objection would indeed be valid if the probabilities for sets of
perceptions in the conscious world were merely propensities for the existence
of these perceptions, so that the particular set of perceptions which are
actualized is not uniquely determined by the theory.  However, Sensible Quantum
Mechanics instead gives the picture of {\it all} sets of perceptions with
nonzero measure as actually occurring.  In this way the theory is really not
fundamentally probabilistic in the propensity sense even for the conscious
world.  Instead, it is a {\it many-perceptions} theory, with the probabilities
to be interpreted almost in the frequency sense as the ratios of numbers of
perceptions that actually exist in the various sets.  (I say ``almost,''
because the probabilities need not be rational ratios, as the ``numbers'' of
perceptions in the sets really refer to the measured continua of perceptions in
the sets.)

 Thus Sensible Quantum Mechanics is very closely related to the ``many-worlds''
interpretation \cite{E,DG}, in which probabilities also have the ``frequency''
interpretation of being the ratios of measures.  In both the quantum state
never collapses as a result of a measurement or perception.  The main
difference is that in Sensible Quantum Mechanics, the ``many'' applies to
conscious perceptions rather than to anything in the unconscious quantum world.

\section{Why I Claim Probabilism Is a Myth}

\hspace{.25in}Because the many-worlds interpretation is not fundamentally
probabilistic in the propensity sense, it seems to give a complete theory for
the quantum world itself that is probabilistic in the frequency sense.  So why
should I claim that such probabilism is a myth?  My argument is that it seems
to be both {\it ugly} and {\it unnecessary} to ascribe probabilities (even in
merely the ``frequency'' sense) to the unconscious quantum world.

 The {\it ugliness} of applying probabilities to the quantum world lies in the
arbitrariness of the choice of {\it which} set of possibilities is to be
assigned probabilities.  This is the uncertainty of which set of amplitudes
should be squared to get probabilities.

 For example, one viewpoint is that it is the amplitude for each
macroscopically distinct outcome of a ``measurement'' that should be squared to
get a probability.  (More precisely, this view is that one takes the
expectation values of a complete set of orthogonal projection operators, each
representing one of the macroscopically-distinct outcomes of the measurement
process.)  This viewpoint has the difficulty or ugliness of requiring the
specification of precisely what a ``measurement'' is supposed to be, and
precisely which projection operators are supposed to be measured by it.

 A more inclusive viewpoint is that the expectation value of {\it any}
projection operator is a probability for the corresponding ``event.''  An even
broader viewpoint is that one can square the amplitude given by projecting the
wavefunction not just by {\it one}, but by a whole {\it sequence} of (possibly
noncommuting) projection operators representing a ``history'' or sequence of
``events.''  (For the resulting probabilities to obey the sum rules under a
coarse-graining of the projection operators, the sequences must obey certain
consistency conditions \cite{Griffiths,Omnes}.)  One can extend this viewpoint,
of assigning probabilities to ``consistent histories,'' yet further to the
viewpoint that one can project the wavefunction by sums of sequences of
projection operators that represent coarser-grained histories.  (Then one needs
``decoherence'' conditions for the resulting probabilities to obey the sum
rules for ``decohering histories''
\cite{GMH,H,DH,I,Hall,IL,ILS,DK,S}.)
%[9-17].)
An even further extension is the viewpoint that probabilities are the real
parts of the expectation values of sums of sequences of projection operators,
whenever these obey a ``linear positivity'' condition of being nonnegative,
giving probabilities for ``linearly positive histories'' that automatically
obey the sum rules \cite{GP}.

 In each of these broader viewpoints there is a family of many different
allowed sets of possibilities (e.g., the family of different complete sets of
orthogonal projection operators in the first viewpoint of the previous
paragraph, or the family of consistent sequences of projection operators in the
second viewpoint, etc.).  To get a set of frequency-type probabilities that sum
to unity, there must apparently be a mysterious choice of a unique set of
possibilities out of the family of all such sets of possibilities.  In the
absence of any definite simple principle for selecting this set, its choice,
and the resulting many-worlds (or perhaps ``many-histories'') theory for the
unconscious quantum world itself, seems ugly.

 One conceivable proposal for allowing probabilism without requiring quite so
much ugliness is the idea that all possibilities in all sets of possibilities
in one of these families actually occur, with measures given by the quantum
expectation values in one of the ways discussed above.  Since, for a normalized
quantum state, these expectation values are designed to add up to unity for a
single set of possibilities, the measures will sum to more than unity when one
adds up all the sets in a family.  In fact, typically the number of allowed
sets of possibilities is infinite.  For example, even for a two-state spin-half
system, there is a complete set of two orthogonal rank-one projection operators
for each direction in space, and hence an infinite number of such sets in the
family of all complete sets of rank-one projection operators.  This means that
the sum of the measures for all sets of possibilities is unnormalizable, which
is problematic.

 This problem may be circumventable for certain choices of the family of sets
of possibilities.  For example, suppose that the family is given by the set of
all decompositions of the identity operator into an ordered set of $m$
orthogonal projection operators $P_i$ of respective ranks $r_i, i=1,\ldots,m,$
for a quantum system with a Hilbert space of dimension $n=\sum_{i=1}^{m}{r_i}$.
 This family forms the proper flag manifold $U(n)/\prod_{i=1}^{m}{U(r_i)}$, a
compact homogeneous manifold of real dimension $n^2-\sum_{i=1}^{m}{r_i^2}$.
For each point on this flag manifold, the expectation values of the
corresponding projectors give a set of $m$ probabilities that sum to unity.  Of
course, just as in the $n=2$ case discussed above, the sum of the expectation
values over the continuous infinity of points of the proper flag manifold
diverges.  However, the proper flag manifold can be given a natural homogeneous
volume element that integrates to unit volume over the entire manifold.  Then
the expectation value of the $i$th projector $P_i$ can be reinterpreted as a
probability {\it density} for the positive outcome of that projector.  If one
integrates this probability density over a finite volume of the proper flag
manifold, one can interpret the result as the joint probability that the set of
possibilities is within that region of the flag manifold and that the $i$th
projector has a positive outcome.

 For a family of consistent or decohering histories, defined by sets of class
operators $C_{\alpha}(x^i)$ that depend upon some parameters or coordinates
$x^i$ that locally label each such set of histories in the family, one could
introduce the Riemannian metric
 \begin{equation}
 g_{ij}dx^idx^j = \sum_{\alpha}
 {Re\,Tr\{[C_\alpha^\dagger(x^i+dx^i)-
 C_\alpha^\dagger(x^i)]
 [C_\alpha(x^j+dx^j)-C_\alpha(x^j)]\}}
 \label{eq:1}
 \end{equation}
and then take the volume element of this metric (normalized by the total volume
over the entire family).  Then one could integrate the expectation value of
$C_\alpha^\dagger(x^i)C_\alpha(x^i)$ (the probability assigned by the
consistent or decohering histories approach to the particular outcome $\alpha$
in the family labeled by the particular coordinates $x^i$, now to be
reinterpreted as a probability density over families of histories) over this
normalized volume element for the families in some set of ranges of the $x^i$'s
to get the joint probability that the set of histories is within that region
and that the outcome $\alpha$ is positive.

 In this way one could get a ``many-many-worlds'' interpretation with
frequency-type probabilities for the unconscious quantum world in which not
only are all {\it possibilities} (with nonzero expectation value) within one
set of possibilities interpreted as actually occurring, but also all {\it sets}
of possibilities within an appropriate family are interpreted as actually
occurring.  However, one might object that even this construction leaves it
ambiguous which {\it family} of sets of possibilities should be chosen.  For
example, I might prefer to choose the family given by the set of all
decompositions of the identity operator into an ordered set of orthogonal
rank-one projection operators, whereas someone else might prefer the family of
all decohering histories.  One might also object that the constructions
outlined above for these two particular classes of families (those decomposing
the identity operator into orthogonal projection operators $P_i$ and those
decomposing it into other suitable operators $C_{\alpha}$) are rather
cumbersome and hence do not completely avoid the problem of ugliness.

 These attempts to ascribe probabilities to the unconscious quantum world
appear not only to be ugly but also to be {\it unnecessary}.  Probabilities
from an accepted theory can help us to predict what types of experiences we may
expect, and the probabilities assigned to our experiences by an uncertain
theory may be used to test that theory, but those predictions and tests
strictly apply to experiences as consciously perceived.  We really have no way
of assigning a meaning to probabilities for things that are not consciously
experienced, and neither can we test such probabilities.  Probabilities can be
arbitrarily assigned to things in the unconscious quantum world, as in the
``many-many-worlds'' interpretation outlined above, but without relating them
to conscious perceptions, their meaning is ambiguous.  Such probabilities are
fundamentally unnecessary for selecting and using a theory for our experiences.

 This diatribe against taking probabilism as a fundamental truth is not meant
to be a denial of the current historical circumstance that for heuristic
purposes it may often be convenient to use the myth of probabilism to assign
``probabilities'' to things in the unconscious quantum world as a substitute or
approximation for what I am arguing are the more fundamental probabilities, in
the ``frequency'' sense, for sets of conscious perceptions.  (It is similarly
often extremely convenient to use the myth of classical physics.)  If such
``probabilities'' in the quantum world are remembered to be purely mythical,
than one can often simply use the ordinary many-worlds interpretation with some
arbitrary choice of the set of possibilities, without worrying about the
arbitrariness of this choice.  A different choice would merely give a different
set of mythical probabilities, useful perhaps for approximating the true
probabilities for some different sets of conscious perceptions.  Only if the
probabilities are to be interpreted as something fundamental need one worry
about the ambiguity or ugliness in the choice in the set of possibilities.

 After writing this section, I heard Coleman's beautiful lecture, ``Quantum
Mechanics in Your Face'' \cite{Sid95}, a rerun of his 1993 Dirac Memorial
Lecture with a more censored title \cite{Sid93}, which argues for ``NO special
measurement process, NO reduction of the wavefunction, NO indeterminacy,
NOTHING probabilistic in quantum mechanics.''  Coleman further states that a
interaction of an infinite set of spins with a measuring device can lead to ``a
definite deterministic state, definitely a random sequence.''
(See the forthcoming \cite{LS} for a proof of this.)
He argues that it is a mistake to call the Everett interpretation
``many worlds'' and prefers that it be called the ``unitarian interpretation''
of quantum mechanics.

 The present paper agrees with this viewpoint for the unconscious quantum
world, though Albert's summary \cite{A92} (p.~124) of Newman's argument against
``the bare theory'' persuades me that we need something like probabilities
(e.g., in the ``frequency'' sense) for perceptions in the conscious world to
explain what is typical about our experience.  For example, if I imagine a
perception of remembering having thrown two million fair coins and knowing how
many came up heads, I would expect a typical such perception to have a memory
of getting within a few thousand of one million heads rather than of getting
merely a few thousand heads total, even though I would expect there to be
nonzero amplitudes for both sets of corresponding brain states.  Unfortunately,
showing that tossing an infinite number of coins may under suitable
circumstances (fair coins, etc.) definitely lead to a random infinite sequence
of heads or tails \cite{LS} does not seem to help explain the experience we
have with finite sequences, since {\it any} finite sequence can be the
beginning of an infinite random sequence.

 One might accept that we do need something beyond bare quantum mechanics to
predict that our actual finite measurement results are typical but object to
appealing to consciousness to do that.  This might be a valid objection if all
we are trying to explain can be described purely without reference to
consciousness.  But the idea of Sensible Quantum Mechanics is not that one
needs to consider consciousness in order to provide a suitable interpretation
of the unconscious quantum world (which I believe is adequately described by
the bare ``unitarian'' theory that Coleman advocates), but rather that one
needs to consider consciousness precisely when one is seeking to explain
properties of conscious experiences themselves.

\section{Axioms of Sensible Quantum Mechanics}

\hspace{.25in}Sensible Quantum Mechanics (SQM) is given by the following three
basic postulates or axioms:

 {\bf Quantum World Axiom}:  The unconscious ``quantum world'' $Q$ is
completely described by an appropriate algebra of operators and by a suitable
state $\sigma$ (a positive linear functional of the operators) giving the
expectation value $\langle O \rangle \equiv \sigma[O]$ of each operator $O$.

 {\bf Conscious World Axiom}:  The ``conscious world'' $M$, the set of all
perceptions $p$, has a fundamental measure $\mu(S)$ for each subset $S$ of $M$.

 {\bf Quantum-Consciousness Connection}:  The measure $\mu(S)$ for each set $S$
of conscious perceptions is given by the expectation value of a corresponding
``awareness operator'' $A(S)$, a positive-operator-valued (POV) measure
\cite{Dav}, in the state $\sigma$ of the quantum world:
 \begin{equation}
 \mu(S) = \langle A(S) \rangle \equiv \sigma[A(S)].
 \label{eq:2}
 \end{equation}

 The Quantum World Axiom is here deliberately vague as to the precise nature of
the algebra of operators and of the state, because as the details of various
quantum theories of the universe are being developed, I do not want the general
framework of Sensible Quantum Mechanics at this time to be made too
restrictive.  For example, SQM is not designed necessarily to exclude the
possibility that the operators may be the pairs of multi-time (or even more
general) class operators (or perhaps even arbitrary linear combinations of
them) in the decohering histories approach
\cite{GMH,H,DH,I,Hall,IL,ILS,DK,S,GP},
%[9-18],
with the expectation values being given by decoherence functionals, even though
SQM does reject as mythical the usual probability interpretation given these
functionals for the unconscious quantum world.

 In the Conscious World Axiom and elsewhere in this paper, a perception $p$ is
taken to mean all that one is consciously aware of or consciously experiencing
at one moment (or, more strictly, the entirety of a single conscious
experience).  Lockwood, in a book expressing remarkably similar ideas to those
that I initially arrived at independently \cite{Lo}, calls $p$ a ``phenomenal
perspective'' or ``maximal experience.''  It could also be expressed as a total
``raw feel'' that one has at once.  It can include components such as a visual
sensation, an auditory sensation, a pain, a conscious memory, a conscious
impression of a thought or belief, etc., but it does not include a sequence of
more than one immediate perception that in other proposals might be considered
to be strung together to form a stream of consciousness of an individual mind.

 I should emphasize that by a perception, I mean the phenomenal, first-person,
subjective experience, and not the processes in the brain (which I call quantum,
even if they are accurately described classically) that accompany these
subjective phenenomena.  In his first chapter, Chalmers \cite{Chal}
gives an excellent discussion of the distinction between the former,
which he calls the phenomenal concept of mind, and the latter,
which he calls the psychological concept of mind.
In his language, what I mean by a perception (and by other
approximate synonyms that I might use, such as sensation
or awareness) is the phenomenal concept, and not the psychological one.

 I should perhaps also emphasize that each perception $p$ has a unique
character given by its content (including its {\it qualia}), so, by definition,
there are no pairs of different perceptions with precisely the same character.
In this way perceptions are different from the interpretation of points of a
connected manifold in general relativity, since there the points are all
equivalent (e.g., under active diffeomorphisms) until one lays down suitably
inhomogeneous fields on the manifold that can then be used to distinguish the
points.  In contrast, for the set $M$ of perceptions, the individual
perceptions $p$ are assumed to be distinguished entirely by their content
before any other structures are laid down, such as the measure $\mu(S)$ for
each subset $S$ of $M$.  The appropriate description of the distinguishing
features of each perception appears to be a difficult problem that I am
generally leaving aside from my discussion of the framework of Sensible Quantum
Mechanics, so in this way my discussion is definitely incomplete.  For my
purposes here, I am merely assuming that such a distinction between all
different perceptions $p$ can in principle be given (though perhaps not in
practice by any conscious being within our universe).

 The Quantum-Consciousness Connection Axiom states my assumption of
the structure of what might be called the `psychophysical laws,'
the laws that presumably give the `neural correlates of consciousness.'
This axiom, when combined with the other two, gives what to me seems to be
the simplest and most conservative framework for ``{\it bridging} principles
that link the physical facts with consciousness''
and for stating ``the connection at the level of
`Brain state X produces conscious state Y'
for a vast collection of complex physical states
and associated experiences'' \cite{Chal}
in language that is consistent with a quantum theory having
``NO special measurement process, NO reduction of the wavefunction,
NO indeterminacy'' \cite{Sid93,Sid95}
(in particular, with a many-perceptions variant of
Everett's quantum theory, in which measures
for sets of conscious perceptions are added
to the bare unitary quantum theory that Coleman advocates).

 Of course, the Quantum-Consciousness Connection Axiom,
like the Quantum world Axiom, is here also deliberately vague,
because I do not have a detailed theory of consciousness,
but only a framework for fitting it with quantum theory.
My suggestion is that any theory of consciousness that is not
inconsistent with bare quantum theory should be formulated
within this framework (unless a better framework can be found, of course).
I am also suspicious of any present detailed theory that purports to
say precisely under what conditions in the quantum world
would consciousness occur, since it seems that we simply don't know yet.
I feel that present detailed theories may be analogous to the cargo cults
of the South Pacific after World War II,
in which an incorrect theory was adopted,
that aircraft with goods would land simply
if airfields and towers were built.

 Since all sets $S$ of perceptions with $\mu(S) > 0$ really occur in the
framework of Sensible Quantum Mechanics,
it is completely deterministic if the quantum state and the $A(S)$
are determined:  there are no random or truly probabilistic elements in SQM.
Nevertheless, because the framework has measures for sets of perceptions,
one can readily use them to calculate quantities
that can be interpreted as conditional probabilities.
One can consider sets of perceptions $S_1$, $S_2$, etc.,
defined in terms of properties of the perceptions.
For example, $S_1$ might be the set of perceptions in
which there is a conscious memory of having tossed a coin one hundred times,
and $S_2$ might be the set of perceptions in which there is a conscious memory
of getting more than seventy heads.  Then one can interpret
 \begin{equation}
 P(S_2|S_1)\equiv \mu(S_1\cap S_2)/\mu(S_1)
 \label{eq:3}
 \end{equation}
as the conditional probability that the perception is in the set $S_2$, given
that it is in the set $S_1$.  In our example, this would be the conditional
probability that a perception included a conscious memory of getting more than
seventy heads, given that it included a conscious memory of having tossed a
coin one hundred times.

 An analogue of this conditional ``probability'' is the conditional probability
that a person in early 1997 is the Queen of England.  If we consider a model of
all the five to six billion people, including the Queen, that we agree to
consider as living humans on Earth in 1997, then at the basic level of this
model the Queen certainly exists in it; there is nothing random or
probabilistic about her existence.  But if the model weights each of the five
to six billion people equally, then one can in a manner of speaking say that
the conditional probability that one of these persons is the Queen is somewhat
less than $2\!\!\times\!\! 10^{-10}$.  I am proposing that it is in the same
manner of speaking that one can assign conditional probabilities to sets of
perceptions, even though there is nothing truly random about them at the basic
level.

 Another analogue one could give for the meaning
of the measures of perceptions postulated in Sensible Quantum Mechanics
(particularly when they are incommensurate) is the following picture,
not to be taken literally, but to be taken as an aid for conceptualizing the measure:
Assume for simplicity that the total measure for all perceptions is finite,
and assume that the number of perceptions is countable.
Then imagine that God moves His finger across each perception,
staying at each one so that it occurs for a ``time''
that is proportional to the measure.
Of course this ``time'' should not be confused with any physical time
within our universe, or with any conscious awareness of time within
any of the perceptions, but it should merely used as a picture
for a continuous variable that can illustrate the measure.
The picture is then that the measure for perceptions may be viewed as
somewhat analogous to the measure of time used for calculating
time averages in dynamical systems, for example.

 As it is defined by the three basic axioms above, Sensible Quantum Mechanics
is a framework and not a complete theory for the universe, since it would need
to be completed by giving the detailed algebra of operators and state of the
quantum world, the set of all possible perceptions  of the conscious world, and
the awareness operators $A(S)$ for the subsets of possible perceptions, whose
quantum expectation values are the measures for these subsets.

 Furthermore, even if such a complete theory were found,
it would not necessarily be the final theory of the universe,
since one would like to systematize the connection
between the elements given above.
As Chalmers eloquently puts it on pages 214-15 of his book \cite{Chal},
``An ultimate theory will not leave the connection at the level of
`Brain state X produces conscious state Y' for a vast collection
of complex physical states and associated experiences.
Instead, it will systematize this connection via an underlying
explanatory framework, specifying simple underlying laws
in virtue of which the connection holds.  Physics does not content
itself with being a mere mass of observations about the positions,
velocities, and charges of various objects at various times;
it systematizes these observations and shows how they are
consequences of underlying laws, where the underlying laws
are as simple and as powerful as possible.
The same should hold of a theory of consciousness.
We should seek to explain the supervenience of consciousness
upon the physical in terms of the simplest possible set of laws.

 ``Ultimately, we will wish for a set of {\it fundamental laws}.
Physicists seek a set of basic laws simple enough that one might
write them on the front of a T-shirt; in a theory of consciousness,
we should expect the same thing.  In both cases, we are questing
for the basic structure of the universe, and we have good reason
to believe that the basic structure has a remarkable simplicity.
The discovery of fundamental laws may be a distant goal, however.
\ldots

 ``When we finally have fundamental theories
of physics and consciousness in hand,
we may have what truly counts as a theory of everything.
The fundamental physical laws will explain the character
of physical processes; the psychophysical laws
will explain the conscious experiences that are associated;
and everything else will be a consequence.''

 Returning to the elements above of a postulated completed,
but not necessarily final, Sensible Quantum Mechanics theory,
it is presently premature to try to give these elements precisely,
particularly the awareness operators
that have generally been left out of physics discussions.
However, it might be helpful to hypothesize certain simplifying forms
that these awareness operators might have.
Before doing even that, it is useful to consider the structure
of the set of all possible perceptions
and to postulate a prior measure for that set.

\section{Hypotheses for a Prior Measure}

\hspace{.25in}I shall hypothesize that the set $M$ of all possible conscious
perceptions $p$ is a suitable topological space with a prior measure that I
shall denote as
 \begin{equation}
 \mu_0(S) = \int_{S}{d\mu_0(p)}.
 \label{eq:4}
 \end{equation}
Then, because of the linearity of positive-valued-operator measures over sets, one can write each awareness operator as
 \begin{equation}
 A(S) = \int_S E(p)d\mu_0(p),
 \label{eq:5}
 \end{equation}
a generalized sum or integral of ``experience operators'' or ``perception operators'' $E(p)$ for the individual perceptions $p$ in the set $S$.  Similarly, one can
write the measure on a set of perceptions $S$ as
 \begin{equation}
 \mu(S) =  \langle A(S) \rangle = \int_S d\mu(p)
   = \int_S m(p) d\mu_0(p),
 \label{eq:6}
 \end{equation}
in terms of a measure density $m(p)$ that is the quantum expectation value of
the experience operator $E(p)$ for the same perception $p$:
 \begin{equation}
 m(p) = \langle E(p) \rangle \equiv \sigma[E(p)].
 \label{eq:7}
 \end{equation}

 Strictly speaking, the prior measure $\mu_0(S)$ is not an essential part of a
complete Sensible Quantum Mechanics theory, since once the algebra of
operators, the quantum state, the set of perceptions, and the awareness
operators are specified, the theory is complete.  However, since we do not yet
have any precise knowledge of these elements, the prior measure is a very
helpful tool to use while postulating and testing various hypotheses about
these elements.

 Perhaps the simplest hypothesis that one can make about the set $M$ of all
possible perceptions is that it is countably discrete.  Such a class of
Sensible Quantum Mechanics theories with discrete perceptions may be labeled
SQMD.  This leads to a natural prior measure
 \begin{equation}
 \mu_{0D}(S) = N(S) = \mbox{number of perceptions in $S$}.
 \label{eq:8}
 \end{equation}
Then the ``integrals'' in Eqs.~(\ref{eq:4})-(\ref{eq:6}) are simply sums over
the discrete perceptions $p$ that make up the set $S$.

 In the alternative hypothesis of SQMC in which the set of all possible
perceptions forms a continuum, there is not such an obvious natural prior
measure.  If the awareness operators are of trace class, one might choose
 \begin{equation}
 \mu_{0CT}(S) = Tr[A(S)],
 \label{eq:9}
 \end{equation}
which is equivalent to requiring $Tr[E(p)]=1$.  This would be reasonable if the
set of possible quantum states formed a finite-dimensional Hilbert space, or
perhaps if each experience operator depended nontrivially upon all of the
quantum system except for a finite-dimensional subsystem, but these
possibilities seems unduly restrictive, so I am doubtful that $\mu_{0CT}(S)$
and the resulting SQMCT is a realistic choice except in toy models.

 Another hypothesis that one might make in the case in which $M$ is a continuum
is to assume that there is a preferred prior state $\sigma_0$ and then to take
the prior measure to be the measure given by the expectation values of the
awareness operators in this state (rather than in the actual state $\sigma$,
which gives the true measure):
 \begin{equation}
 \mu_{0P}(S) = \sigma_0[A(S)].
 \label{eq:10}
 \end{equation}
Such a subset of the SQMC theories might be labeled SQMC$\sigma_0$.  If the set
of possible quantum states forms a Hilbert space of finite dimension $n$, one
might choose $\sigma_0$ to be the maximally mixed state, the one giving
$\sigma_0[O] = Tr[O]/n$ for any operator $O$.  Then $\mu_{0P}(S)$ would simply
be $\mu_{0CT}(S)/n$.  However, $\mu_{0P}(S)$ could be defined in more general
situations in which $\mu_{0CT}(S)$ is not definable, though in such situations
there might not be such a natural choice for $\sigma_0$.  It may be hard enough
to find a simple state $\sigma$ for the quantum world of our universe that fits
observations, without the added difficulty of requiring also the specification
of a prior state $\sigma_0$.  (In quantum field theory in Minkowski spacetime,
the Minkowski vacuum would appear to be a natural choice for $\sigma_0$, but
finding a state $\sigma$ in that framework consistent with our observations
seems nontrivial.  When one includes gravitation, the situation seems reversed
in that there may be a certain natural choices for $\sigma$
\cite{Haw,HarHaw,Vil,Lin,LLM},
%[25-29],
but then one seems to lose the Minkowski vacuum as a natural choice for
$\sigma_0$.)

 Yet another hypothesis one might make for the prior measure $\sigma_0$ in the
continuous case is that it is given by the Riemannian volume element
 \begin{equation}
 \mu_{0R}(S) = \int_S g^{1/2}(p) d^np
 \label{eq:11}
 \end{equation}
of a Riemannian metric on the set of set $M$ of all perceptions.  If one took
the set of experience operators $E(p)$ as basic rather than the awareness
operators $A(S)$ to be derived from them by Eq.~(\ref{eq:5}), one might define
the Riemannian metric to be
 \begin{equation}
 g_{ij}dp^idp^j = Re\,Tr\{[E(p^i+dp^i)-E(p^i)][E(p^j+dp^j)-E(p^j)]\}
\label{eq:12}
 \end{equation}
if this is both finite and nondegenerate.  In such a case both the prior
measure $\mu_{0R}(S)$ and the awareness operators $A(S)$ would follow from a
single suitable set of experience operators $E(p)$.  One might label such a
family of Sensible Quantum Mechanics theories SQMCR, the final R denoting the
Riemannian metric (\ref{eq:12}) and the corresponding prior volume element
$\mu_{0R}(S)$.

\section{Hypotheses for Experience Operators}

\hspace{.25in}After one has chosen a suitable prior volume element $\mu_0(S)$
so that one then can derive the awareness operators $A(S)$ from the experience
operators $E(p)$, one can formulate various hypotheses for the latter.
Clearly, one could get any measure density $m(p)$ one wanted from
Eq.~(\ref{eq:7}) applied to {\it any} arbitrary quantum state $\sigma$ simply
by choosing $E(p)=m(p)I$, where $I$ is the identity operator (so long as the
state $\sigma$ satisfies the normalization requirement $\sigma[I] \equiv
\langle I \rangle = 1$).  However, this would put no burden of the explanation
of the measure of perceptions onto the quantum state.  I strongly suspect that
this would not lead to the simplest possible choice of $E(p)$'s giving the
correct measure density $m(p)$, since presumably the actual quantum state is a
crucial element in the simplest complete description of even just the conscious
world.  (Otherwise, since one's perception in the conscious world is all one
can be aware of, why should one even postulate the existence of the quantum
world?  Surely the justification we have for the postulate of the existence of
the quantum world is that such a postulate simplifies the explanation of the
conscious world that we directly experience.  In this way the whole physical
world, comprising both the conscious world and the quantum world, related by
the Quantum-Consciousness Connection axiom, can be simpler than just the
conscious world considered by itself.  Whether the whole physical world is also
simpler than just the quantum world considered by itself is a question posed
for the future; the fact that it does not seem to be so within our present
level of understanding seems to me to be one of the main reasons why
consciousness has so far been largely banished from physics.)

 One very natural requirement that we may wish to put on the experience
operators is the following:

 {\bf Pairwise Independence Hypothesis}:  $E(p)$ and $E(p')$ are linearly
independent for two different perceptions, $p\neq p'$.

 This would be sufficient to rule out the ad hoc proposal $E(p)=m(p)I$ above in
which the quantum state would have no effect on the measure for perceptions.
Of course, it is much stronger than necessary for merely that purpose.  It
implies that no two different perceptions have the same experience operator (or
even the same up to normalization).  The Quantum-Consciousness Connection and
the assumption of a prior measure that leads to Eq.~(\ref{eq:5}) says that
there is a unique map from the set of perceptions $M$ in the conscious world to
a subset of operators in the quantum world that I am calling experience
operators $E(p)$; the Pairwise Independence Hypothesis says that there is a
unique inverse map from this subset of operators back to the set of
perceptions.  These assumptions together provide a particular type of mind-body
unity that seems rather plausible (or at least relatively simple), though it
certainly is not logically necessary, and I doubt that there could even be any
direct observational evidence for it apart from appeals to simplicity.

 Unless explicitly stated otherwise, in the following I shall generally assume
the Pairwise Independence Hypothesis.  Where necessary, Sensible Quantum
Mechanics with this additional assumption may be denoted SQMI (or SQMDI in the
discrete case, etc., skipping the P for ``pairwise'' to avoid confusion with
the P below for ``projection''), though I shall generally just assume the
Pairwise Independence Hypothesis implicitly and not bother listing the I for
it.

 A related but much stronger hypothesis than the Pairwise Independence
Hypothesis is the following:

 {\bf Linear Independence Hypothesis}:  The set of {\it all} the $E(p)$'s is a
linearly independent set.  Such SQM theories may be labeled SQMLI.

 If the $E(p)$'s are positive operators in a Hilbert space of finite complex
dimension $N$, then there are at most $N^2$ linearly independent such
operators, so that would be the limit on the number of perceptions $p$ in SQMLI
in a finite-dimensional Hilbert space.  This may be a plausible restriction,
but since it does not seem to be necessary, and since I do not see too much
other motivation for the Linear Independence Hypothesis (except as a possible
consequence of other hypotheses that I may wish to consider), I shall not
generally assume it but shall only implicitly assume the Pairwise Independence
Hypothesis below when I speak in an unqualified way about Sensible Quantum
Mechanics.

 Now one can turn to more particular assumptions about the structure of the
experience operators.  One of the simplest hypotheses one can make about them
is that they are projection operators:

 {\bf Projection Hypothesis}:  $E(p)=P(p)$, a projection operator that depends
on the perception $p$.

 Such forms of Sensible Quantum Mechanics can be denoted by attaching the
letter P to the abbreviation.  Thus the general form of SQM with the Projection
Hypothesis added is SQMP (or SMQIP if it is necessary to show explicitly that
the projection operators are all assumed to be different for different
perceptions); the form with a discrete set of perceptions (each with its own
projection operator) would be SQMDP (or SQMDIP); the form with a continuum of
perceptions with the prior measure given by the volume element of the
Riemannian metric (\ref{eq:12}) would be SQMCRP (or SQMCRIP); etc.

 The Projection Hypothesis appears to be a specific mathematical realization of
part of Lockwood's proposal \cite{Lo} (p. 215), that ``a phenomenal perspective
[what I have here been calling simply a perception $p$] may be equated with a
shared eigenstate of some preferred (by consciousness) set of compatible brain
observables.''  Here I have expressed the ``equating'' by Eqs.~(\ref{eq:2}) and
(\ref{eq:5}), and presumably the ``shared eigenstate'' can be expressed by a
corresponding projection operator $P(p)$.

 I should also emphasize that if the same conscious perception is produced by
several different orthogonal ``eigenstates of consciousness'' (e.g., different
states of a brain and surroundings that give rise to the same perception $p$),
then in the Projection Hypothesis the projection operator $P(p)$ would be a sum
of the corresponding rank-one projection operators and so would be a projection
operator of rank higher than unity.  This is what I would expect, since surely
the surroundings could be different and yet the appropriate part of the brain,
if unchanged, would lead to the same perception.  On the other hand, if $E(p)$
were a sum of noncommuting projection operators corresponding to nonorthogonal
states, or a weighted sum of projection operators with weights different from
unity, then generically $E(p)$ would not be a projection operator $P(p)$ as
assumed in the Projection Hypothesis.

 If one has a constrained system, such as a closed universe in general
relativity, the quantum state may obey certain constraint equations, such as
the Wheeler-DeWitt equations.  The projection operators $P(p)$ of perceptions
in the Projection Hypothesis may not commute with these constraints, in which
case they may give technically `unphysical' states when applied to the quantum
state.  But so long as their expectation values can be calculated and are
nonnegative real numbers, that is sufficient for giving the perception measure
density $m(p)$.  What it means is that in the Projection Hypothesis, the
perception operators should be considered as projection operators in the space
of unconstrained states, even though the actual physical state does obey the
constraints.

 Alternatively, if one wishes to write the perception operators $E(p)$ as
operators within the space of constrained states, the Projection Hypothesis
could be modified to the following assumption to give perception operators
$E(p)$ that commute with the constraints and so keep the state `physical':

 {\bf Constrained Projection Hypothesis}:  $E(p)=P_CP(p)P_C$, where $P_C$ is a
projection operator within the space of unconstrained states that takes any
state to a corresponding constrained state, and $P(p)$ is a projection operator
in the space of unconstrained states that depends on the perception $p$.

 One might label such theories SQMP(C), SQMDP(C), SQMCRP(C), etc., where the C
for ``constrained'' is put in parentheses to indicate that it modifies the P
and to distinguish it from a possible earlier C for ``continuum'' or a later
one for ``commuting'' (see below).

 One can also get something like the Constrained Projection Hypothesis, say the
{\bf Symmetrized Projection Hypothesis}, even for unconstrained systems if they
have symmetries (e.g., the Poincar\'{e} symmetries of quantum field theory in a
classical Minkowski spacetime, though one would not expect these symmetries to
survive when one includes gravity), since one might then expect that $E(p)$
should be invariant under the symmetry group with elements $g$.  Then if one
starts with a projection operator $P(p)$ that is not invariant under the action
of each group element, say $P(p)\neq gP(p)g^{-1}$, then one might expect $E(p)$
to be proportional to the sum or integral of $gP(p)g^{-1}$ over the group
elements $g$, so that $E(p) = gE(p)g^{-1}$.  Unless all these different
$gP(p)g^{-1}$'s are orthogonal (which does not appear possible for a continuous
symmetry group), the resulting $E(p)$ will generically not be a projection
operator, but it can be said to have arisen from one.  Such theories might be
labeled SQMP(S), SQMDP(S), SQMCRP(S), etc.

 One might prefer to make an even more restrictive assumption and strengthen
the Projection Hypothesis to the Linearly Independent Projection Hypothesis,
the Commuting Projection Hypothesis or, stronger yet, the Orthogonal Projection
Hypothesis:

 {\bf Linearly Independent Projection Hypothesis}:  $E(p)=P(p)$, with the set
of {\it all} these $P(p)$'s being a linearly independent set.  (This
combination of the Projection Hypothesis with the Linear Independence
Hypothesis is much stronger than the combination of the Projection Hypothesis
with the now-implicit Pairwise Independence Hypothesis given earlier above,
which merely requires that each {\it pair} of different $E(p)$'s be linearly
independent.)  Such SQM theories may be labeled SQMLIP.

 {\bf Commuting Projection Hypothesis}:  $E(p)=P(p)$, with $[P(p),P(p')]=0$ for
all pairs of perceptions $p$ and $p'$.  Such SQM theories may be labeled SQMPC,
putting the C for ``commuting'' after the P for ``projection.''

 {\bf Orthogonal Projection Hypothesis}:  $E(p)=P(p)$, with $P(p)P(p')=0$ for
all pairs of {\it different} perceptions $p \neq p'$.  Such SQM theories (a
subset of the SQMPC theories) may analogously be labeled SQMPO.  Unlike the
Projection Hypothesis and the Commuting Projection Hypothesis by themselves,
the Orthogonal Projection Hypothesis obviously implies the Pairwise
Independence Hypothesis and the Linear Independence Hypothesis.

 The Commuting Projection Hypothesis and the Orthogonal Projection Hypothesis
allow one to define measures on {\it pairs} of perceptions that have nice
properties.  However, such joint measures are not fundamental to SQM and may be
viewed as analogous to yet another case of mythical probabilities.  Therefore,
the possibility of their definition in SQMPC and SQMPO theories does not seem
to me to be a strong argument in favor of such restriction of SQMP theories.

 Another direction one can go in hypothesizing properties of experience
operators is to make the following assumption about the structure of each
perception (which, as defined above, is all that one is aware of at once, or
all of a single conscious experience):

 {\bf Assumption of Perception Components}:  Each perception $p$ itself
consists of a set of discrete components $c_i(p)$ contained within the
perception, say $p = \{c_i(p)\}$.  Different perceptions $p$ may share various
components in common, but at least some components must differ in order that
the perceptions themselves differ.

 Because of the apparent unity of perception, the Assumption of Perception
Components seems likely to be more of an approximation for certain aspects of
perceptions than a general truth.  In other words, there may not be any
fundamental decomposition of perceptions into components.  However, there are
cases in which it appears to be a reasonably good approximation, such as for a
perception with one component being a conscious memory of having tossed a coin
one hundred times, and with another component being a conscious memory of
getting more than seventy heads.

 If one does make the Assumption of Perception Components, perhaps just as an
approximation for certain aspects of a perception that are easy to describe, it
may be natural to make the following extension of the Projection Hypothesis:

 {\bf Product Projection Hypothesis}:  $E(p)=\prod_{i}{P[c_i(p)]}$, where each
$P[c_i(p)]$ is a projection operator that depends on the perception component
$c_i(p)$, with all the $P[c_i(p)]$'s commuting for the {\it same} $p$.  The
corresponding theories, subsets of those obeying the Projection Hypothesis, can
be labeled by a double PP replacing the single P.

 One can also strengthen the Product Projection Hypothesis in ways analogous to
those in which the Projection Hypothesis were strengthened above:

 {\bf Commuting Product Projection Hypothesis}:  $E(p)=\prod_{i}{P[c_i(p)]}$,
where each $P[c_i(p)]$ is a projection operator that depends on the perception
component $c_i(p)$, with {\it all} the $P[c_i(p)]$'s commuting (i.e., for all
{\it different} $p$'s, as well as for all $c_i(p)$'s with the same $p$ as in
the Product Projection Hypothesis itself).  The resulting theories may be
labeled SQMPPC.

 {\bf Orthogonal Product Projection Hypothesis}:
$E(p)=P(p)=\prod_{i}{P[c_i(p)]}$, with all the $P[c_i(p)]$'s commuting, and
with $P(p)P(p')=0$ for all pairs of {\it different} perceptions $p \neq p'$.
In other words, each pair of different perceptions has at least one
corresponding pair of different components whose corresponding projection
operators are orthogonal.  Like the Orthogonal Projection Hypothesis, the
Orthogonal Product Projection Hypothesis implies both the Pairwise Independence
Hypothesis and the Linear Independence Hypothesis.  Such theories may be
labeled SQMPPO.

 After discussing all of these stronger hypotheses that one might add to the
Projection Hypothesis (or to one of its slight variants, such as the
Constrained Projection Hypothesis or the Symmetrized Projection Hypothesis), I
should say that one might prefer to stop at an even weaker hypothesis, such as
the following:

 {\bf Sequence of Projections Hypothesis}:  $E(p)=C^{\dagger}(p)C(p)$, where
\\$C(p)=P(p,n)P(p,n-1)\cdots P(p,2)P(p,1)$ is a product of a sequence of
(possibly noncommuting) projection operators (or a ``homogeneous history''),
with the integer $n$ and the projection operators $P(p,i)$ all depending on the
perception $p$.  Such theories can be labeled by adding S to the previous
abbreviation (as P is added for theories obeying the Projection Hypothesis).
Note that when $n=1$, SQMS reduces to SQMP.

 For $n>1$, one might prefer to restrict the sequences to those which obeys the
conditions of ``consistent histories'' \cite{Griffiths,Omnes}.  Thus one might
make the following restriction:

 {\bf Consistent Sequence of Projections Hypothesis}:  The Sequence of
Projections Hypothesis holds, and the set of all experience operators $E(p)$
forms a subset of a single set of consistent histories, that is, a set of
histories with the property that, for each pair of distinct sequences $C(p)$
and $C(p')$ such that $\hat{C}(p,p')=C(p)+C(p')$ is also a homogeneous history
(a product of a sequence of projection operators),
$\sigma[\hat{C}(p,p')^{\dagger}\hat{C}(p,p')] = \sigma[C^{\dagger}(p)C(p)] +
\sigma[C^{\dagger}(p')C(p')]$.  Such a theory can be labeled SQMSC.

 An example obeying the Sequence of Projections Hypothesis would be the case in
which the individual projection operators are those of the components $c_i(p)$
of the perception $p$.  However, if these projection operators are actually
noncommuting, contrary to the Product Projection Hypothesis above, it may not
be clear what fixes the order of the sequence.  For example, it would not seem
natural to try to impose any time ordering, since the perception is all that
one is aware of ``at once'' and so might be interpreted as happening at one
time.  One might try to order the components of a perception that are
interpreted as memories by the order in the time at which the events remembered
are felt to have happened, but this seems very imprecise in many cases and
certainly would not seem to apply to the components of a perception that are
felt to be present sensations of concurrent events.

 {\bf Histories Hypothesis}:  $E(p)=C(p)^{\dagger}C(p)$, where each $C(p)$ is a
(possibly one-term) {\it sum} (with unit coefficients) of products of sequences
of projection operators.  Such theories can be labeled by adding H instead of P
or S to the previous abbreviation.  SQMH is a further generalization of SQMP
than SQMS is; SQMP is a subset of SQMS, which is itself a subset of SQMH.

 Just as one might like to restrict the sequences of projection operators in
SQMS to consistent histories, so one might analogously want to restrict the
$C$'s in SQMH to one or another of the ``decoherent histories'' requirements
\cite{GMH,H,DH,I,Hall,IL,ILS,DK,S,GP}:
%[9-18]:

 {\bf Individually Weak Decoherent Histories Hypothesis}:  The Histories
Hypothesis holds, with each $C(p)$ being a member of a weak decoherent set of
histories consisting of $C(p)$ and $I-C(p)$.  That is,
$\sigma[C(p)^{\dagger}C(p)]=Re\,\sigma[C(p)]$.  Such theories may be called
SQMHIWD.

 {\bf Individually Medium Decoherent Histories Hypothesis}:  The Histories
Hypothesis holds, with each $C(p)$ being a member of a medium decoherent set of
histories consisting of $C(p)$ and $I-C(p)$.  That is,
$\sigma[C(p)^{\dagger}C(p)]=\sigma[C(p)]$.  This version of SQMH may be called
SQMHIMD.

 {\bf Individually Strong Decoherent Histories Hypothesis}:  The Histories
Hypothesis holds, with each $C(p)$ being a member of a strong decoherent set of
histories consisting of $C(p)$ and $I-C(p)$.  That is, there exists a
projection operator $P(p)$ such that $\sigma[OC(p)]=\sigma[OP(p)]$ for any
operator $O$.  This is SQMHISD.

 {\bf Weak Decoherent Histories Hypothesis}:  The Histories Hypothesis holds,
with the $C(p)$'s forming part of a weak decoherent set of histories.  That is,
\\$Re\,\sigma[C(p)^{\dagger}C(p')]=0$ for $p\neq p'$.  This would be SQMHWD.

 {\bf Medium Decoherent Histories Hypothesis}:  The Histories Hypothesis holds,
with the $C(p)$'s forming part of a medium decoherent set of histories.    That
is, $\sigma[C(p)^{\dagger}C(p')]=0$ for $p\neq p'$.  This can be labeled
SQMHMD.

 {\bf Strong Decoherent Histories Hypothesis}:  The Histories Hypothesis holds,
with the $C(p)$'s forming part of a strong decoherent set of histories.  That
is, there exists a set of orthogonal projection operators $P(p)$, a distinct
one for each distinct perception $p$, such that $\sigma[OC(p)]=\sigma[OP(p)]$
for any operator $O$.  This may be denoted SQMHSD.

 In addition to these various forms of the Histories Hypothesis, one can also
consider the extension to Linearly Positive Histories \cite{GP}:

 {\bf Linearly Positive Histories Hypothesis}:  $E(p)=Re\,C(p)$, where each
$C(p)$ is a sum of one or more products of sequences of projection operators
such that $\sigma[Re\,C(p)]\geq 0$.  Such theories, SQMLPH, unlike the previous
versions of the Histories Hypothesis, are not necessarily subsets of SQMH.
However, they give the same measure densities $m(p)$ as SQMHIWD or its subsets
SQMHIMD, SQMHISD, SQMHWD, SQMHMD, or SQMHSD when the $C(p)$'s obey the
corresponding more restrictive conditions of those theories \cite{GP}.

 It is certainly logically possible that perceptions might depend on histories
(characterized by $C$'s that are sums of products of sequences of generically
noncommuting projection operators) rather than on events (characterized by
individual projection operators) that one could consider localized on
hypersurfaces of constant time if the quantum world has such a time.  However,
as a previous advocate of the ``marvelous moment'' approach to quantum
mechanics in which only quantities on one such hypersurface can be tested
\cite{P}, I find it more believable to assume that perceptions are caused by
brain states which could be at one moment of time if there are such things in
the physical world.  The generalization of this hypothesis to the case in which
there may not be a well-defined physical time leads me personally to prefer
adopting the Projection Hypothesis (or perhaps the Constrained Projection
Hypothesis for constrained systems) rather than stopping at the Sequence of
Projections Hypothesis or the Histories Hypothesis, though of course the
general framework of Sensible Quantum Mechanics is broad enough to encompass
any of these more specific hypotheses.

 On the other hand, I am not myself convinced that the evidence strongly
suggests going beyond the Projection Hypothesis to the Commuting Projection
Hypothesis, the Orthogonal Projection Hypothesis, the Product Projection
Hypothesis, or any of the restrictions of the latter, though it would certainly
be worth investigating these more specific hypotheses to see whether one of
them might indeed lead to a simple complete theory that is consistent with our
experience.  Some of the properties implied by these various hypotheses will be
discussed in more detail in the section below on toy models.

 While I am considering hypotheses weaker than the Projection Hypothesis, it
might be worth listing a few other very weak requirements, restricting only the
normalization of the experience operators $E(p)$, that nevertheless would
generally be sufficient to exclude the ad hoc proposal $E(p)=m(p)I$:

 {\bf Constant-Maximum-Normalization Hypothesis}:  The expectation value of
each $E(p)$ has a constant maximum value, say unity, in the set of all states
$\sigma$ that are normalized to give $\sigma[I]=1$.  This weaker hypothesis
would be a consequence of the Projection Hypothesis but of course does not
imply the Projection Hypothesis.  Such theories could be called SQMNCM, putting
the initial for ``normalization'' before that of its modifiers.

 {\bf Unit-Normalization Hypothesis}:  $Tr[E(p)]=1$ for each experience
operator.  This is a consequence of Eq.~(\ref{eq:9}) for SQMCT or SQMDT, but it
is not consistent with the Projection Hypothesis unless all the projection
operators $P(p)$ are of rank one, which seems extremely implausible.  These
theories would be SQMNU.

 {\bf Projection-Normalization Hypothesis}:  Each experience operator is
normalized so that $Tr[E(p)]=Tr[E(p)E(p)]$.  This is a consequence (at least if
$E(p)$ is of trace class so that the left side of this normalization condition
is defined) of the Projection Hypothesis, since then $E(p)=E(p)E(p)$.  Such
theories might be labeled SQMNP.  However, the trace-class condition on $E(p)$
that seems necessary for imposing the Projection-Normalization Hypothesis seems
unrealistic to impose if the quantum system has an infinite number of states.

 Finally, for comparison's sake, let me describe some hypotheses that would
take one outside Sensible Quantum Mechanics itself, as I have defined it by the
three axioms above, by violating the particular Quantum-Consciousness
Connection axiom given by Eq.~(\ref{eq:2}).  These broader alternative
hypotheses would give the measure $\mu(S)$ on sets $S$ of perceptions as {\it
nonlinear} functionals of the quantum state.  For example, one might set
 \begin{equation}
 \mu(S) = \int_S f[p,m(p)] d\mu_0(p)
 \equiv  \int_S f(p,\langle E(p) \rangle) d\mu_0(p)
 \equiv  \int_S f(p,\sigma[E(p)]) d\mu_0(p),
 \label{eq:12b}
 \end{equation}
where $f$ is some arbitrary nonnegative finite function, depending possibly
upon the perception $p$ itself, of its other argument, the nonnegative number
$m(p)$.  The simplest class would be those in which the function is purely of
$m(p)$, i.e., $f[p,m(p)]=f[m(p)]$.  A simple set of examples would be to set
$f[m(p)]=m(p)^n$ for some positive constant $n$.  Such extensions of Sensible
Quantum Mechanics might be labeled SQMf, or SQM$n$ in the case in which $f$ is
the power-law function.  Of course, SQM1, the case in which the power is $n=1$,
is ordinary Sensible Quantum Mechanics, which is what I shall henceforth assume
unless explicitly stated otherwise.

 One can see that there are many possible characteristics for the general form
of the experience operators in Sensible Quantum Mechanics or its extensions
SQMf.  From one viewpoint this merely illustrates the incompleteness of the
bare framework of SQM, but in a slightly different way of looking at it, it
shows part of the enormous gap of our knowledge about consciousness, even
within this one framework, which itself is merely a proposal.

\section{Sensible Classical Mechanics}

\hspace{.25in}For illustrative purposes, it may be helpful to note that one
could have a rather similar relation between consciousness and mechanics even
if mechanics were classical.  Then one might propose a theory of {\it Sensible
Classical Mechanics} (SCM) with the following three axioms analogous to those
proposed above for Sensible Quantum Mechanics:

 {\bf Classical World Axiom}:  The unconscious ``classical world'' $C$ is
completely described by an appropriate set of classical histories and by a
particular history $h$ within that set.

 {\bf Conscious World Axiom}:  The ``conscious world'' $M$, the set of all
perceptions $p$, has a fundamental measure $\mu(S)$ for each subset $S$ of $M$.

 {\bf Classical-Consciousness Connection}:  The measure $\mu(S)$ for each set
$S$ of conscious perceptions is given by the value of a corresponding
``classical awareness functional'' $A_C(S)$, a positive functional, linear in
the set $S$ of histories, evaluated for the specific history $h$ of the
classical world:
 \begin{equation}
 \mu(S) = A_C(S)[h].
 \label{eq:12c}
 \end{equation}

 Just as in SQM, so also in SCM one might postulate that the set $M$ of
perceptions $p$ forms a suitable topological space with a prior measure given
by Eq.~(\ref{eq:4}) and then set
 \begin{equation}
 A_C(S) = \int_S E_C(p)d\mu_0(p),
 \label{eq:12d}
 \end{equation}
a generalized sum or integral of ``classical experience operators'' or
``classical perception operators'' $E_C(p)$ for the individual perceptions $p$.
 Then one could use the relevant part of Eq.~(\ref{eq:6}) to define the measure
density $m(p)$ as
 \begin{equation}
 m(p) = E_C(p)[h],
 \label{eq:12e}
 \end{equation}
the analogue of Eq.~(\ref{eq:7}) above.

 In some forms of Sensible Classical Mechanics, one might take the set of
classical histories to be time-parametrized trajectories, obeying some set of
classical equations of motion, in some phase space.  Then one might take the
value of a classical awareness functional $A_C(S)[h]$ to be a time integral,
along the trajectory of the actual history $h$ in the phase space, of some
positive function of the phase space that is linear in $S$.  For example,
$E_C(p)[h]$ could be the time integral, along the history $h$, of a suitable
delta-function of the phase space, so that each perception has a particular
corresponding point of the phase space and so that the measure density $m(p)$
is, say, unity if the trajectory of the actual history $h$ goes through the
point in the phase space corresponding to the perception $p$ and is zero
otherwise.

 Then if the set of all perceptions were homeomorphic to some subset of the
phase space, and if the correspondence between each perception and its
corresponding point of this subset of the phase space were a homeomorphism,
then an actual classical history passing through the subset of the phase space
would give a corresponding one-parameter trajectory though the set of all
perceptions.  This is indeed what one might na\"{\i}vely expect in a classical
model for a single conscious being having a continuous temporal sequence of
perceptions, but even within SCM one could certainly have many different
variants of this simple example.  (For example, one would probably want to
allow for more than one conscious being, so that appropriate points of the
phase space would correspond to several conscious perceptions, say one for each
conscious being that existed at that point of the phase space.)  However, I
shall not bother listing various detailed possibilities as I did above for the
more realistic possibility of Sensible Quantum Mechanics.

 Sensible Classical Mechanics seems rather moot, since the unconscious
``mechanical'' part of the physical world appears to be quantum rather
classical.  However, I give it mainly to show that the connection I am
proposing between the mechanical and the conscious aspects of the world is not
strongly dependent upon the quantum nature of mechanics, though the detailed
mathematical form of the connection does depend upon the mathematical form of
the description of the mechanics.   (Contrast \cite{St} for an opposing
viewpoint in which the form of the mechanics is considered crucial.)  In other
words, if it turns out  that a classical approximation is adequate for the
mechanics of the brain that leads to consciousness, then there would not seem
to be any fundamental problem with using a particular SCM as an approximation
to the appropriate SQM.  For example, if SQM is correct and the experience
operators $E(p)$ are approximately projection operators onto certain brain
configurations that can be described classically to a good approximation, then
a SCM with perceptions corresponding to the regions of phase space with those
brain configurations would presumably be a good approximation to at least that
part of the SQM in which the quantum state behaves according to the classical
approximation for it.

 Another use for Sensible Classical Mechanics might be in attaching a Sensible
theory of conscious perceptions to the de Broglie-Bohm theory of quantum
mechanics
\cite{deB,Bohm,BH,Be,DGZ,A92,Ho,A94,BDDGZ,Page}
%[32-36, 22, 37-39]
in which there is not only the algebra of operators and the quantum state, but
also a classical history or de Broglie-Bohm trajectory (albeit not one obeying
the same equations of motion as that of the classical approximation to the
quantum theory).  If SQM is applied to this theory according to the rules
above, then the measure for conscious perceptions would be completely
unaffected by the de Broglie-Bohm trajectory, so that this trajectory would
have absolutely no influence upon what is consciously experienced.  (Thus my
advocacy of SQM leaves me with no personal motivation for augmenting quantum
mechanics with what would then be a totally unobservable de Broglie-Bohm
trajectory.)

 However, someone else who does believe in the existence of a de Broglie-Bohm
classical trajectory and who does believe it has an observable effect might
thus choose not to adopt SQM but might prefer to adopt some form of SCM with
the history given by the de Broglie-Bohm trajectory (giving, say, an SCMBB
theory).  Although I myself would not advocate doing that, it seems that one
must have something similar in mind if one believes that a de Broglie-Bohm
version of quantum mechanics is in principle observationally different from
quantum mechanics without the de Broglie-Bohm trajectories.  Nevertheless, the
statistical predictions for perceptions in this SCMBB theory could be similar
to those of certain SQM theories in which the experience operators are
projection operators in configuration space, at least if one averages over a
suitable prior distribution for the unknown precise de Broglie-Bohm trajectory
of the former.  Thus the observational difference that would be manifest if one
had access to the entire conscious world could very well not be sufficient to
make the two theories distinguishable by any single typical perception.
Indeed, many advocates of a de Broglie-Bohm version of quantum mechanics would
say that it is observationally indistinguishable from ordinary quantum
mechanics.  (See the third paper of \cite{Page} for a further discussion of these points.)  

 After this interlude on how relating consciousness to classical mechanics need
not be all that different from relating it to quantum mechanics, I shall return
to the assumption that bare quantum mechanics is the correct framework for the
mechanical aspects of the physical world and that Sensible Quantum Mechanics is
the correct framework for the combination of the mechanical and the conscious
aspects.

\section{Perceptions rather than Minds}

\hspace{.25in}Another point I should emphasize is that in Sensible Quantum
Mechanics, the set $M$ of perceptions is fundamental, but not any higher power
of this set.  Thus there is a linear measure on subsets $S$ of perceptions,
which can be expressed as the ``integral'' (\ref{eq:6}) (a discrete sum when
the set $M$ is discrete) of a measure density $m(p)$ times a prior measure
element $d\mu_0(p)$, but there is no nontrivial fundamental measure density
$m(p_1,p_2,\ldots,p_n)$ on $n$-tuples of perceptions.  Thus, for example, there
is no fundamental notion of a correlation between individual perceptions given
by any measure.

 (On the other hand, if a perception can be broken up into component parts, say
$c_1$ and $c_2$, there can be a correlation between the parts, in the sense
that the measure $\mu(S_1\cap S_2)$ for all perceptions in the set $S_1$
containing the component $c_1$ and in the set $S_2$ containing the component
$c_2$ need not be proportional to $\mu(S_1)\mu(S_2)$, the measure for all
perceptions containing $c_1$ times the measure for all perceptions containing
$c_2$.  The enormous structure in a single perception seems to suggest that
such correlations within perceptions are highly nontrivial, but I see no
evidence for ascribing any fundamental meaning to a nontrivial correlation
between complete perceptions $p$, since no two different complete perceptions
can be perceived together.)

 In saying that SQM posits no fundamental correlation between complete
perceptions, I do not mean that it is impossible to define such correlations
from the mathematics, but only that I do not see any fundamental physical
meaning for such mathematically-defined correlations.  As an example of how
such a correlation might be defined, consider that if a perception operator
$E(p)$ is a projection operator, and the quantum state of the universe is
represented by the pure state $|\psi\rangle$, one can ascribe to the perception
$p$ the pure Everett ``relative state''
 \begin{equation}
 |p\rangle=\frac{E(p)|\psi\rangle}{\parallel  E(p)|\psi\rangle\parallel}
 =\frac{E(p)|\psi\rangle}
 {\langle\psi|E(p)E(p)|\psi\rangle^{1/2}}.
 \label{eq:P3}
 \end{equation}
Alternatively, if the quantum state of the universe is represented by the
density matrix $\rho$, one can associate the perception with a relative density
matrix
 \begin{equation}
 \rho_p=\frac{E(p)\rho E(p)}{Tr[E(p)\rho E(p)]}.
 \label{eq:P4}
 \end{equation}
Either of these formulas can be applied when the perception operator is not a
projection operator, but then the meaning is not necessarily so clear.

 Then if one is willing to say that $m(p)=Tr[E(p)\rho]$ is the absolute
probability for the perception $p$ (which might seem natural at least when
$E(p)$ is a projection operator, though I am certainly not advocating this
na\"{\i}ve interpretation), one might also na\"{\i}vely interpret
$Tr[E(p')\rho_p]$ as the conditional probability of the perception $p'$ given
the perception $p$.

 Another thing one can do with two perceptions $p$ and $p'$ is to calculate an
``overlap fraction'' between them as
 \begin{equation}
 f(p,p')=\frac{\langle E(p)E(p')\rangle\langle  E(p')E(p)\rangle}
 {\langle E(p)E(p)\rangle\langle E(p')E(p')\rangle}.
 \label{eq:P5}
 \end{equation}
If the quantum state of the universe is pure, this is the same as the overlap
probability between the two Everett relative states corresponding to the
perceptions:  $f(p,p')=|\langle p|p'\rangle|^2$.  Thus one might in some sense
say that if $f(p,p')$ is near unity, the two perceptions are in nearly the same
one of the Everett ``many worlds,'' but if $f(p,p')$ is near zero, the two
perceptions are in nearly orthogonal different worlds.  However, this is just a
manner of speaking, since I do not wish to say that the quantum state of the
universe is really divided up into many different worlds.  In a slightly
different way of putting it, one might also propose that $f(p,p')$, instead of
$Tr[E(p')\rho_p]$, be interpreted as the conditional probability of the
perception $p'$ given the perception $p$.  Still, I do not see any evidence
that $f(p,p')$ should be interpreted as a fundamental element of Sensible
Quantum Mechanics.  In any case, one can be conscious only of a single
perception at once, so there is no way in principle that one can test any
properties of joint perceptions such as $f(p,p')$.

 An amusing property of both of these ``conditional probabilities'' for one
perception given another is that they would both always be zero if the
Orthogonal Projection Hypothesis were true.  Even though the resulting SQMPO
theory would generally be a ``many-perceptions'' theory, it could be
interpreted as being rather solipsistic in the sense that in the relative
density matrix $\rho_p$ corresponding to my present perception, no other
perceptions would occur in it with nonzero measure!  This has the appearance of
being somewhat unpalatable, and might be taken to be an argument against
adopting the Orthogonal Projection Hypothesis (and hence perhaps for stopping
at the Commuting Projection Hypothesis, or perhaps the Commuting Product
Projection Hypothesis if one adopts the Assumption of Perception Components, as
the strongest reasonable hypothesis), but it is not clear to me that this is
actually strong evidence against the Orthogonal Projection Hypothesis.

 In addition to the fact that Sensible Quantum Mechanics postulates no
fundamental notion of any {\it correlation} between individual perceptions, it
also postulates no fundamental {\it equivalence} relation on the set of
perceptions.  For example, the measure gives no way of classifying different
perceptions as to whether they belong to the same conscious being (e.g., at
different times) or to different conscious beings.  The most reasonable such
classification would seem to be by the content (including the {\it qualia}) of
the perceptions themselves, which distinguish the perceptions, so that no two
different perceptions, $p\neq p'$, have the same content.  Based upon my own
present perception, I find it natural to suppose that perceptions that could be
put into the classification of being alert human perceptions have such enormous
structure that they could easily distinguish between all of the $10^{11}$ or so
persons that are typically assigned to our history of the human race.  In other
words, in practice, different people can presumably be distinguished by their
conscious feelings.

 Another classification of perceptions might be given by classifying the
perceptions operators $E(p)$ rather than the content of the perceptions
themselves.  This would be more analogous to classifying people by the quantum
nature of their bodies (in particular, presumably by the characteristics of
their brains).  However, I doubt that in a fundamental sense there is any
absolute classification that uniquely distinguishes each person in all
circumstances.  (Of course, one could presumably raise this criticism about the
classification of any physical object, such as a ``chair'' or even a
``proton'':  precisely what projection operators correspond to the existence of
a ``chair'' or of a ``proton''?)  Therefore, in the present framework
perceptions are fundamental, but persons (or individual minds), like other
physical objects, are not, although they certainly do seem to be very good
approximate entities (perhaps as good as chairs or even protons) that I do not
wish to deny.  Even if there is no absolute definition of persons in the
framework of Sensible Quantum Mechanics itself, the concept of persons and
minds does occur in some sense as part of the {\it content} of my present
perception, just the concepts of chairs and of protons do (in what are perhaps
slightly different ``present perceptions,'' since I am not quite sure that I
can be consciously aware of all three concepts at once, though I seem to be
aware that I have been thinking of three concepts).

 In this way the framework of Sensible Quantum Mechanics proposed here is a
particular manifestation of Hume's ideas \cite{Hume}, that ``what we call a
{\it mind}, is nothing but a heap or collection of different perceptions,
united together by certain relations, and suppos'd, tho' falsely, to be endow'd
with a perfect simplicity and identity'' (p. 207), and that the self is
``nothing but a bundle or collection of different perceptions'' (p. 252).  As
he explains in the Appendix (p. 634), ``When I turn my reflexion on {\it
myself}, I never can perceive this {\it self} without some one or more
perceptions; nor can I ever perceive any thing but the perceptions.  'Tis the
composition of these, therefore, which forms the self.''  (Here I should note
that what Hume calls a perception may be only one {\it component} of the
``phenomenal perspective'' or ``maximal experience'' \cite{Lo} that I have been
calling a perception $p$, so one $p$ can include ``one or more perceptions''
$c_i(p)$ in Hume's sense.)

 Furthermore, each experience or perception operator $E(p)$ need not have any
precise location in either space or time associated with it, so there need be
no fundamental place or time connected with each perception.  Indeed, Sensible
Quantum Mechanics can easily survive a replacement of spacetime with some other
structure (e.g., superstrings) as more basic in the quantum world.  Of course,
the {\it contents} of a perception can include a sense or impression of the
time of the perception, just as my present perception when I perceive that I am
writing this includes a feeling that it is now A.D. 1995, so the set of
perceptions $p$ must include perceptions with such beliefs, but there need not
be any precise time in the physical world associated with a perception.  That
is, perceptions are `outside' physical spacetime (even if spacetime is a
fundamental element of the physical world, which I doubt).

 As a consequence of these considerations, there are no unique time-sequences
of perceptions to form an individual mind or self in Sensible Quantum
Mechanics.  In this way the present framework appears to differ from those
proposed by Squires \cite{Sq}, Albert and Loewer \cite{A,A92}, and Stapp
\cite{St}.  (Stapp's also differs in having the wavefunction collapse at each
``Heisenberg actual event,'' whereas the other two agree with mine in having a
fixed quantum state, in the Heisenberg picture, which never collapses.)
Lockwood's proposal \cite{Lo} seems to be more similar to mine, though he also
proposes (p. 232) ``a continuous infinity of parallel such streams'' of
consciousness, ``{\it differentiating} over time,'' whereas Sensible Quantum
Mechanics has no such stream as fundamental.  On the other hand, later Lockwood
\cite{Lo2} does explicitly repudiate the Albert-Loewer many-minds
interpretation, so there seems to me to be little disagreement between
Lockwood's view and Sensible Quantum Mechanics except for the detailed
formalism and manner of presentation.  Thus one might label Sensible Quantum
Mechanics as the Hume-Everett-Lockwood-Page (HELP) interpretation, though I do
not wish to imply that these other three scholars, on whose work my proposal is
heavily based, would necessarily agree with my present formulation.

 Of course, the perceptions themselves can include components that  seem to be
memories of past perceptions or events.  In this way it can be a very good
approximation to give an approximate order for perceptions whose content
include memories that are correlated with the contents of other perceptions.
It might indeed be that the measure density $m(p)$ for perceptions including
detailed memories is rather heavily peaked around approximate sequences
constructed in this way.  But I would doubt that the contents of the
perceptions $p$, the perception operators $E(p)$, or the measure densities
$m(p)$ for the set of perceptions would give unique sequences of perceptions
that one could rigorously identify with individual minds.

 Because the physical state of our universe seems to obey the second law of
thermodynamics, with growing correlations in some sense, I suspect that the
measure density $m(p)$ may have rather a smeared peak (or better, ridge) along
approximately tree-like structures of branching sequences of perceptions, with
perceptions further out along the branches having contents that includes
memories that are correlated with the present-sensation components of
perceptions further back toward the trunks of the trees.  This is different
from what one might expect from a classical model with a discrete number of
conscious beings, each of which might be expected to have a unique sharp
sequence or non-branching trajectory of perceptions.  In the quantum case, I
would expect that what are crudely viewed as quantum choices would cause
smeared-out trajectories to branch into larger numbers of smeared-out
trajectories with the progression of what we call time.  If each smeared-out
trajectory is viewed as a different individual mind, we do get roughly a
``many-minds'' picture that is analogous to the ``many-worlds'' interpretation
\cite{E,DG}, but in my framework of Sensible Quantum Mechanics, the ``many
minds'' are only approximate and are not fundamental as they are in the
proposal of Albert and Loewer \cite{A}.  Instead, Sensible Quantum Mechanics is
a ``many-perceptions'' or ``many-sensations'' interpretation.  One might thus
label it philosophically as Mindless Sensationalism.

 Even in a classical model, if there is one perception for each conscious being
at each moment of time in which the being is conscious, the fact that there may
be many conscious beings, and many conscious moments, can be said to lead to a
``many-perceptions'' interpretation.  However, in Sensible Quantum Mechanics,
there may be vastly more perceptions, since they are not limited to a discrete
set of one-parameter sharp sequences of perceptions, but occur for all
perceptions $p$ for which $m(p)$ is positive.  In this way a quantum model may
be said to be even ``more sensible'' (or is it ``more sensational''?) than a
classical model.  One might distinguish SQM from a classical model like SCM
with many perceptions by calling SQM a ``very-many-perceptions'' framework,
meaning that almost all (say as defined by the prior measure) possible
perceptions actually occur with nonzero measure density.  (Thus SQM might, in a
narrowly literal sense, almost be a version of panpsychism, but the enormous
range possible for the logarithm of the measure density means that it is really
quite far from the usual connotations ascribed to panpsychism.  This is perhaps
comparable to noting that there may be a nonzero amplitude that almost any
system, such as a star, has a PC in it, and then calling the resulting
many-worlds theory pancomputerism.)

 One might fear that the present attack on the assumption of any definite
notion of a precise identity for persons or minds as sequences of perceptions
would threaten human dignity.  Although I would not deny that I feel that it
might, I can point out that on the other hand, the acceptance of the viewpoint
of Sensible Quantum Mechanics might increase one's sense of identity with all
other humans and other conscious beings.  Furthermore, it might tend to
undercut the motivations toward selfishness that I perceive in myself if I
could realize in a deeply psychological way that what I normally anticipate as
my own future perceptions are in no fundamental way picked out from the set of
all perceptions.  (Of course, what I normally think of as my own future
perceptions are presumably those that contain memory components that are
correlated with the content of my present perception, but I do not see
logically why I should be much more concerned about trying to make such
perceptions happy than about trying to make perceptions happy that do not have
such memories:  better to do unto others as I would wish they would do unto
me.)  One can find that Parfit \cite{Par} had earlier drawn similar, but much
more sophisticated, conclusions from a view in which a unique personal identity
is not fundamental.

\section{Quantum Field Theory Model}

\hspace{.25in}Although Sensible Quantum Mechanics transcends quantum theories
in which space and time are fundamental, and although I believe that such
theories will need to be transcended to give a good theory of our universe, it
might help to get a better feel for the spacetime properties of perceptions by
considering the context of quantum field theory in an unquantized curved
globally-hyperbolic background spacetime in which spacetime points are
unambiguously distinguished by the spacetime geometry (so that the Poincar\'{e}
symmetries are entirely broken and one need not worry about integrating over
$gP(p)g^{-1}$'s to satisfy superselection rules for energy, momentum, and/or
angular momentum \cite{PW}).  This simplified model might in some sense be a
good approximation for part of the entire quantum state of the universe in a
correct theory if there is one that does fit into the framework of Sensible
Quantum Mechanics and does give a suitable classical spacetime approximation.

 In the Heisenberg picture used in this paper, the quantum state is independent
of time (i.e., of a choice of Cauchy hypersurface in the spacetime), but the
Heisenberg equations of evolution for the fundamental fields and their
conjugate momenta can be used to express the operators $E(p)$ in terms of the
fields and momenta on any Cauchy hypersurface.  The arbitrariness of the
hypersurface means that even in this quantum field theory with a well-defined
classical spacetime, and even with a definite foliation of the spacetime by a
one-parameter (time) sequence of Cauchy hypersurfaces, there is no unique
physical time that one can assign to any of the perceptions $p$; they are
`outside' time as well as space.

 Furthermore, the operators $E(p)$ in this simplified model are all likely to
be highly nonlocal in terms of local field operators on any Cauchy
hypersurface, since quantum field theories that we presently know do not seem
to have enough local operators to describe the complexities of an individual
perception, unless one considers high spatial derivatives of the field and
conjugate momentum operators.  However, for a given one-parameter (time)
sequence of Cauchy hypersurfaces, one might rather arbitrarily choose to define
a preferred time for each perception $p$ as the time giving the Cauchy
hypersurface on which the corresponding $E(p)$, if expressed in terms of fields
and momenta on that hypersurface, has in some sense the smallest spatial spread
at that time.

 For example, to give a tediously explicit {\it ad hoc} prescription, on a
Cauchy hypersurface labeled by the time $t$ one might choose a point $P$ and a
ball that is the set of all points within a certain geodesic radius $r$ of the
point.  Then one can define an operator $E'(p;t,P,r)$ that is obtained from
$E(p)$ written in terms of the fields and conjugate momenta at points on the
hypersurface by throwing away all contributions that have any fields or
conjugate momenta at points outside the ball of radius $r$ from the point $P$.
Now define the overlap fraction
 \begin{equation}
 F(p;t,P,r)=\frac{\langle E(p)E'(p;t,P,r)\rangle
 \langle E'(p;t,P,r)E(p)\rangle}
 {\langle E(p)E(p)\rangle
 \langle E'(p;t,P,r)E'(p;t,P,r)\rangle}.
 \label{eq:P6}
 \end{equation}
(If both  $E(p)$ and $E'(p;t,P,r)$ were projection operators, and the actual
quantum state were a pure state, then $F$ would be the overlap probability
between the states obtained by projecting the actual quantum state by these
projectors and normalizing.)  If $E(p)$ is nonlocal, this fraction $F$ will be
small if the radius $r$ is small but will be nearly unity if the radius $r$ is
large enough for the ball to encompass almost all of the Cauchy hypersurface.
For each perception $p$, time $t$, and point $P$, one can find the smallest $r$
that gives $F=1/2$, say, and call that value of the radius $r(p;t,P)$.  Then
one can find the point $P=P(p;t)$ on the hypersurface that gives the smallest
$r(p;t,P)$ on that hypersurface for the fixed perception $p$ and call the
resulting radius $r(p;t)$.  Finally, define the preferred time $t_p$ as the
time $t$ for which $r(p;t)$ is the smallest, and label that smallest value of
$r(p;t)$ for the fixed perception $p$ as $r_p$.

 If the perception operator $E(p)$ for some human conscious perception is not
unduly nonlocal in the simplified model under present consideration, and if the
quantum state of the fields in the spacetime has macroscopic structures that at
the time $t_p$ of the perception are fairly well localized (e.g., with quantum
uncertainties less than a millimeter, say, which would certainly not be a
generic state, even among states which give a significant $m(p)$ for the
perception in question), one might expect that at this time the ball within
radius $r_p$ of the point $P(p;t_p)$ on the hypersurface labeled by $t_p$ would
be inside a human brain.  It would be interesting if one could learn where the
point $P(p;t_p)$ is in a human brain, and what the radius $r_p$ is, for various
human perceptions, and how the location and size of this ball depends on the
perception $p$.

\section{Testing and Comparing Sensible Quantum
Mechanics Theories}

\hspace{.25in}Any proposed theory should be tested against experience before
being accepted.  If one has a theory in which only a small subset of the set of
all possible perceptions is predicted to occur (e.g., a classical theory in
which there are a finite number, one for each conscious being, of
time-sequences of perceptions that are determined by the trajectory in the
phase space of the system), one can simply check whether an experienced
perception is in that subset.  If it is not, that is clear evidence against the
theory.

 The situation is unfortunately more complicated in very-many-perceptions
theories, such as Sensible Quantum Mechanics, in which almost all perceptions
are predicted to occur with nonzero measure density $m(p)$.  Unless one
experienced a perception in the set, say $S_0$, for which the particular SQM
theory under investigation predicts $m(p)=0$, one could not absolutely rule out
that theory.  For a typical SQM theory, the set $S_0$ is of measure zero,
either using the prior measure, or even using the measure of an alternative
typical SQM theory, such as, presumably, the (unknown) correct theory.  I.e.,
it is likely that $\mu'(S_0)=0$ from the measure $\mu'(S)$ for almost any other
theory, such as the unknown correct one.  Thus one is not at all likely to have
a perception that would absolutely rule out almost any specific SQM theory.

 The best one can hope for with a very-many-perceptions theory is to find {\it
likelihood} evidence for or against it, where the likelihood is the probability
that the theory assigns or predicts for an experienced outcome.  Even this
cannot be done directly for a particular experienced perception $p$ in SQM
theories, since they merely assign a measure density $m(p)$ to the perception
and not a probability to it.  One somehow needs to get an assigned probability
for the perception or some aspect of it from the theory.

 In the case of SQMD with a countably discrete set of perceptions, and in the
case in which the total measure $\mu(M)$ of the set $M$ of all perceptions is
finite, one could assign the normalized probability $P_r(p)=m(p)/\mu(M)$ to
each individual perception $p$.  However, even in this highly restricted case,
a very low $P_r(p)$ assigned to the experienced perception would not
necessarily be strong evidence against the particular SQMD theory that made
this assignment, for it might simply be that there are a huge number of
possible perceptions in the theory, each of which is assigned a similarly low
$P_r(p)$.

 If there were only a finite total number $N$ of possible perceptions in some
SQMD theory, one could say that a typical perception $p$ should have $P_r(p)$
not too much lower than $P_{r0}(p) \equiv \mu_0(p)/\mu_0(M)=1/N$, the
probability that one gets from the prior measure that weights each discrete
perception equally.  Thus an experience of a perception $p$ for which the
theory predicts $NP_r(p)\ll 1$ would be evidence against that theory.

 However, the apparent complexity of my present perception suggests to me that
if one restricted attention to theories having only a finite number $N$ of
possible perceptions, the simplest of these theories giving my perception among
the $N$ would have $N$ very large.  (One could logically have a theory in which
only my present perception existed, an utterly extreme form of solipsism, but I
strongly doubt that such a theory could be so simple as a theory with many
possible perceptions.  This is analogous to saying that although logically
there exists a theory which gives the single positive integer
\\8568193572865287529475652899568765824569287623819923752927591010,\\
this theory is not so simple in some sense as the simplest theory which gives
all the first $9^{99}$ positive integers.)  Furthermore, if the simplest such
theory with a finite set of possible perceptions, including mine, gave $N$ very
large, I would suspect that an even simpler theory existed which includes my
perception and in which the total number of possible perceptions is not
restricted to be finite.  (This is analogous to saying that the infinite set of
all positive integers is simpler than the large but finite set of the first
$9^{99}$ integers.)

 Because of such examples showing how an infinite set can easily be simpler
than a finite set, I suspect that the total number of possible perceptions is
infinite.  (An extension of this reasoning further suggests to me that it may
be simpler to have the set of possible perceptions continuous rather than
discrete, though in this case it is less compelling.  If one continued
accepting a sequence of such arguments, one would apparently be led to a set
with infinite cardinality, which seems as if it might in the end be more
complex, but that might just be the appearance to our limited way of thinking.)

 In any case, it would be nice to have a way of calculating the likelihood for
one's perception in SQM theories more general than those with a finite number
of possible perceptions.  To do this, it seems that one needs to go beyond the
probability for the mere existence of the perception (which in the existential
sense is unity in SQM for all perceptions with positive measure density, and
which in the frequency sense is useful only if $\mu(p)/\mu(M)$ is nonzero,
which requires that the perceptions be discrete so that the measure $\mu(p)$
for a single perception be positive, and furthermore requires that the total
measure $\mu(M)$ for the set $M$ of all perceptions be finite).

 One probability assigned by a particular SQM theory to a perception that may
be calculated simply from the measure and the measure density is the
probability that a perception is as ``typical'' as it is, where the (ordinary)
{\it typicality} $T(p)$ of one's perception $p$ may be defined in the following
way if the total measure $\mu(M)$ for all perceptions is finite:  Let
$S_{\leq}(p)$ be the set of perceptions $p'$ with $m(p') \leq m(p)$.  Then
 \begin{equation}
 T(p) \equiv \mu(S_{\leq}(p))/\mu(M).
 \label{eq:13}
 \end{equation}
For $p$ fixed and $\tilde{p}$ chosen randomly with the infinitesimal measure
$d\mu(\tilde{p})$, the probability that $T(\tilde{p})$ is less than or equal to
$T(p)$ is
 \begin{equation}
 P_T(p) \equiv P(T(\tilde{p})\leq T(p)) = T(p).
 \label{eq:14}
 \end{equation}
In the case in which $m(p)$ varies continuously in such a way that $T(p)$ also
varies continuously, this typicality $T(p)$ has a uniform probability
distribution between 0 and 1, but if there is a nonzero measure of perceptions
with the same value of $m(p)$, then $T(p)$ has discrete jumps.  (In the extreme
case in which $m(p)$ is one constant value over all perceptions, $T(p)$ is
unity for each $p$.)

 Using this particular criterion of typicality and assuming that one's
perception $p$ is indeed typical in this regard, one might say that agreement
with observation requires that the prediction, by the theory in question, of
$P_T(p)=T(p)$ for one's observation $p$ be not too much smaller than unity.

 Once one defines a typicality, such as by Eq.~(\ref{eq:13}), one can use a
Bayesian approach and assign prior probabilities $P(H_i)$ to individual
hypotheses $H_i$.  Suppose that each such hypothesis gives a particular SQM
theory in detail, and hence its predictions for a measure density $m_i(p)$ over
all perceptions, from which one can assign a particular typicality $T_i(p)$ to
the perception one experiences.  Then the probability $P_{T_i}(p)$ that the
theory predicts that a random perception would have a typicality no greater
than $T_i(p)$, which by Eq.~(\ref{eq:14}) is $T_i(p)$ itself, may be taken to
be the likelihood of $H_i$ given $p$.  By Bayes' rule, the posterior
conditional probability that one should then rationally assign to the
hypothesis $H_i$, if one followed this prescription of interpreting the
typicality as the conditional probability (given the hypothesis $H_i$) for
one's particular perception $p$, would be
 \begin{equation}
 P(H_i|p)=\frac{P(H_i)T_i(p)}{\sum_{j}^{}{P(H_j)T_j(p)}}.
 \label{eq:15}
 \end{equation}

 The main new difficulty in this Bayesian approach, even if it is assumed that
one can indeed calculate the typicality for the perception given each
hypothesis $H_i$, is the assignment of the prior probabilities $P(H_i)$.  These
probabilities are certainly not the frequency-type probabilities that occur
within one SQM theory, nor are they even ``probabilities'' assigned to the
unconscious quantum world in what I am claiming are some sort of mythical
idealization or approximation for what I am proposing are the true
frequency-type probabilities for the conscious world.  Instead, these prior
probabilities would be purely subjective probabilities (in a way that I am
claiming that the ratios of measures in the conscious world are not; the latter
are supposed to be entirely objective frequency-type probabilities, though
their precise values would be inaccessible to us unless we were given the
correct precise SQM theory for our universe).  The prior probabilities are more
like propensities in that they might be interpreted as our guesses for the
propensities for God to have created a universe according to the particular SQM
theories in question.  (It is conceivable that they could be interpreted as
frequencies for an ensemble of universes described by the various different SQM
theories, but this would require a meta-theory for such an ensemble, which
definitely goes beyond the ensemble of worlds in the Everett many-worlds
interpretation of one single closed quantum system such as the universe, or the
ensemble of conscious perceptions within one single SQM theory and
corresponding physical world.)

 Since the prior probabilities assigned to particular SQM hypotheses $H_i$ are
subjective, they may be assigned rather arbitrarily.  Based upon the goal of
getting a simple complete theory for the universe, one might prefer to choose
them so that simpler theories would be given higher prior probabilities.  For
example, one simple choice for a countably infinite set of hypotheses is the
set of prior probabilities
 \begin{equation}
 P(H_i)=2^{-n_i},
 \label{eq:16}
 \end{equation}
where $n_i$ is the rank of $H_i$ in order of increasing complexity.

 Unfortunately, even to make this simple assignment, one would need to assume
some particular background knowledge with respect to which one might define
``complexity.''  In the cases in which one simply wants to compare the
complexity of very complex items, the background knowledge, if sufficiently
small compared with the information in the items, is not too important.  But
for the goal of finding a complete theory of the universe which may not have a
large amount of information in it, the background knowledge is relatively
important and seems to thwart an attempt to use the simple formula
Eq.~(\ref{eq:16}).  However, since Eq.~(\ref{eq:16}) is a subjective (if
apparently simple) choice anyway, one could simply use it with a subjective
choice of the background knowledge with respect to which the rank of $H_i$ in
order of increasing complexity is made.  The only difficulty is that if a
different choice were made, then even for the same perception $p$, the same set
of hypotheses $H_i$, and the same calculations of the typicalities $T_i(p)$,
Eq.~(\ref{eq:15}) would give a different assignment of the posterior
probabilities for the hypotheses.  It is then an open question whether the
theory that is thus assigned the highest posterior probability would then the
same in both cases, though that could be true if the typicalities assigned by
the various theories varied so much that the posterior probabilities
(\ref{eq:15}) are then relatively stable with respect to the changes in the
prior probabilities (\ref{eq:16}).  (This is indeed roughly what seems to occur
for many well-established theories within present physics, such as Maxwell's
electromagnetism, which most physicists accept within a certain domain of
applicability, since they would assign very low prior probabilities to more
complicated alternatives that fit the data, even if they did not agree
precisely how low to set such prior probabilities.)

 There is also the potential technical problem that one might assign nonzero
prior probabilities to hypotheses $H_i$ in which the total measure $\mu(M)$ for
all perceptions is {\it not} finite, so that the right side of
Eq.~(\ref{eq:13}) may have both numerator and denominator infinite, which makes
the typicality $T_i(p)$ inherently ambiguous.  To avoid this problem, one might
use, instead of $T_i(p)$ in Eq.~(\ref{eq:15}), rather
 \begin{equation}
 T_i(p;S) = \mu_i(S_{\leq}(p)\cap S)/\mu_i(S)
 \label{eq:17}
 \end{equation}
for some set of perceptions $S$ containing $p$ that has $\mu_i(S)$ finite for
each hypothesis $H_i$.  This is related to a practical limitation anyway, since one could presumably only hope to be able to compare the measure densities $m(p)$ for some small set of perceptions rather similar to one's own, though it is not clear in quantum cosmological theories that allow an infinite amount of inflation how to get a finite measure even for a small set of perceptions.  Unfortunately, even if one can get a finite measure by suitably restricting the set $S$, this makes the resulting $P(H_i|p;S)$ depend on this chosen $S$ as well as on the other postulated quantities such as $P(H_i)$.

 Instead of using the particular typicality defined by Eq.~(\ref{eq:13}) above,
one could of course instead use any other property of perceptions which places
them into an ordered set to define a corresponding ``typicality.''  For
example, I might be tempted to order them according to their complexity, if
that could be well defined.  Thinking about this alternative ``typicality''
leaves me surprised that my own present perception seems to be highly
complicated but apparently not infinitely so.  What simple complete theory
could make a typical perception have a high but not infinite complexity?

 However, the ``typicality'' defined by Eq.~(\ref{eq:13}) has the merit of
being defined purely from the prior and fundamental measures, with no added
concepts such as complexity that would need to be defined.  The necessity of
being able to rank perceptions, say by their measure density, in order to
calculate a typicality, is indeed one of my main motivations for postulating a
prior measure Eq.~(\ref{eq:4}).

 Nevertheless, there are alternative typicalities that one can define purely
from the prior and fundamental measures.  For example, one might define a {\it
reversed typicality} $T_r(p)$ in the following way (again assuming that the
total measure $\mu(M)$ for all perceptions is finite):  Let $S_{\geq}(p)$ be
the set of perceptions $p'$ with $m(p') \geq m(p)$.  Then
 \begin{equation}
 T_r(p) \equiv \mu(S_{\geq}(p))/\mu(M).
 \label{eq:13r}
 \end{equation}
For $p$ fixed and $\tilde{p}$ chosen randomly with the infinitesimal measure
$d\mu(\tilde{p})$, the probability that $T_r(\tilde{p})$ is less than $T_r(p)$
is
 \begin{equation}
 P_{T_r}(p) \equiv P(T_r(\tilde{p})\leq T_r(p)) = T_r(p),
 \label{eq:14r}
 \end{equation}
the analogue of Eq.~(\ref{eq:14}) for the ordinary typicality.

 In the generic continuum case in which $m(p)$ varies continuously in such a
way that there is only an infinitesimally small measure of perceptions whose
$m(p)$ are infinitesimally near any fixed value, the reversed typicality
$T_r(p)$ is simply one minus the ordinary typicality, i.e., $1-T(p)$, and also
has a uniform probability distribution between 0 and 1.  Its use arises from
the fact that just as a perception with very low ordinary typicality $T(p)\ll
1$ could be considered unusual, so a perception with an ordinary typicality too
near one (and hence a reversed typicality too near zero, $T_r(p)\ll 1$) could
also be considered unusual, ``too good to be true.''

 Perhaps one might like to combine the ordinary typicality with the reversed
typicality to say that a perception giving either typicality too near zero
would be evidence against the theory.  For example, one might define the {\it
dual typicality} $T_d(p)$ as the probability that a random perception
$\tilde{p}$ has the lesser of its ordinary and its reversed typicalities less
than or equal to that of the perception under consideration:
 \begin{equation}
 T_d(p) \equiv P(\min{[T(\tilde{p}),T_r(\tilde{p})]}\leq  \min{[T(p),T_r(p)}])
\equiv \mu(S_d(p))/\mu(M),
 \label{eq:14d}
 \end{equation}
where $S_d(p)$ is the set of all perceptions $\tilde{p}$ with the minimum of
its ordinary and reversed typicalities less than or equal to that of the
perception $p$, i.e., the set with
$\min{[T(\tilde{p}),T_r(\tilde{p})]}\leq\min{[T(p),T_r(p)]}$.  In the case in
which $T(p)$, and hence also $T_r(p)$, varies continuously from 0 to 1,
 \begin{equation}
 T_d(p) = 1 - | 1 - 2T(p) |.
 \label{eq:14e}
 \end{equation}
Then the dual typicality $T_d(p)$ would be very small if the ordinary
typicality $T(p)$ were very near either 0 or 1.

 Of course, one could go on with an indefinitely long sequence of typicalities,
say making a perception ``atypical'' if $T(p)$ were very near any number of
particular values at or between the endpoint values 0 and 1.  But these
endpoint values are the only ones that seem especially relevant, and so it
would seem rather {\it ad hoc} to define ``typicalities'' based on any other
values.  Since $T_d(p)$ is symmetrically defined in terms of both endpoints
(or, more precisely, in terms of both the $\leq$ and the $\geq$ relations for
$m(p')$ in comparison with $m(p)$), in some sense it seems the most natural one
to use.  Obviously, one could use it, or its modification along the lines of
Eq.~(\ref{eq:17}), instead of $T(p)$ in the Bayesian Eq.~(\ref{eq:15}).

 To illustrate how one might use these typicalities in a Bayesian analysis,
consider the SQM$n$ alteration of SQM given by Eq.~(\ref{eq:12b}) with
$f[m(p)]=m(p)^n$.  Suppose that the exponent $n$ is postulated to be uncertain
(unlike in pure SQM, where it is postulated to be precisely 1), say with a
prior probability distribution $P(n)dn$ simply equal to $dn$ and hence uniform
over all $n$.  This prior distribution is obviously not normalizable, but the
normalization or lack thereof will cancel out in Eq.~(\ref{eq:15}) in the use I
am giving it.

 Now consider a simple toy model in which the perceptions $p$ form a continuum
of one dimension (labeled by a single real number, which for simplicity I shall
also call $p$), with the uniform prior measure $d\mu_0(p)=dp$, and with $m(p)$,
the expectation value of $E(p)$, having a gaussian distribution in $p$, say
$m(p)=e^{-p^2/2}$ with the origin and scale of the numbers $p$ adjusted so that
the mean of the distribution is at $p=0$ and the standard deviation is 1.  Now
suppose that a perception of a particular value $p$ occurs.

 Since $f[m(p)]=m(p)^n=e^{-np^2/2}$ is also a gaussian in $p$ centered at zero,
but with standard deviation $1/\sqrt{n}$, one can readily calculate that for
positive $n$ the typicality of the perception p in the theory labeled by the
exponent $n$ is
 \begin{equation}
 T(p) ={\rm erfc}(\sqrt{n p^2/2})\equiv
 1-{\rm erf}(\sqrt{n p^2/2})\equiv
 1-{2 \over \sqrt{\pi}}
 \int_{0}^{\sqrt{n p^2/2}}{e^{-x^2}dx},
 \label{eq:14f}
 \end{equation}
the reversed typicality is
 \begin{equation}
 T_r(p) =1-T(p)={\rm erf}(\sqrt{n p^2/2}),
 \label{eq:14g}
 \end{equation}
and the dual typicality is
 \begin{eqnarray}
 T_d(p) = 1 - | 1 - 2T(p) | = 1 - | 1 - 2T_r(p) |
 \nonumber \\=1 - | 1 - 2 {\rm erfc} (\sqrt{n p^2/2}) |
 = 1 - | 1 - 2 {\rm erf} (\sqrt{n p^2/2}) |.
 \label{eq:14h}
 \end{eqnarray}
For negative $n$ the measure density $f[m(p)]=m(p)^n$ diverges for large $p$,
so the typicality and dual typicalities are both zero in that case, whereas the
reversed typicality is unity.

 Now we can insert these typicalities and the {\it ad hoc} prior measure
$P(n)dn = dn$ into the Bayesian Eq.~(\ref{eq:15}) to calculate the posterior
probability distribution for the exponent $n$ given a particular perception
$p$.  Using the ordinary typicality $T(p)$, the sum in the denominator of
Eq.~(\ref{eq:15}) becomes an integral over $n$, which can be restricted to
positive $n$, since $T(p)$ vanishes for negative $n$, and which gives the value
$1/p^2$.  Thus the corresponding posterior probability distribution becomes
 \begin{equation}
 P(n|p)dn = p^2{\rm erfc}(\sqrt{p^2 n/2})dn
 \label{eq:14i}
 \end{equation}
for positive $n$, and 0 for negative $n$.

 This probability density $P(n|p)$ is monotonically decreasing with $n$, with
the $m$th moment of $n$ being
 \begin{equation}
 \langle n^m \rangle = {(2m+1)!! \over (m+1)p^{2m}}.
 \label{eq:14ii}
 \end{equation}
(Here and in the remainder of this Section the angular brackets $\langle
\rangle$ denote the expectation value in the probability distribution for $n$
given $p$, not the quantum expectation value in the state $\sigma$ that the
angular brackets denote in other parts of this paper.)  Thus the mean posterior
value for $n$ is $3/(2p^2)=1.5/p^2$, and its standard deviation is
$\sqrt{11}/(2p^2)\approx 1.658312/p^2$.  (The mean and standard deviation are
both larger than 1 even if $p=1$, essentially because we started with the
uniform prior distribution $P(n)dn=dn$ which has an infinite mean and standard
deviation, at least if it is restricted to $n\geq 0$ where the typicality of a
finite $p$ is not 0.  One would get a mean closer to unity if one had instead
started with a prior distribution such as $P(n)dn=[\pi(1+n^2)]^{-1}dn$, which
is invariant under $n\rightarrow 1/n$, but I shall not further consider here
such a more complicated prior distribution.)

 For $n\gg 1/p^2$ the probability distribution of Eq.~(\ref{eq:14i}) has an
exponentially decreasing asymptotic form
 \begin{equation}
 P(n|p)dn \sim \sqrt{{2\over \pi p^2 n}}
 e^{-{1\over 2}p^2n}.
 \label{eq:14j}
 \end{equation}
Thus a perception of, say, $p\sim 1$, which is roughly what one would expect if
the exponent $n$ were indeed 1 as SQM would give, would by Eq.~(\ref{eq:15}),
with the ordinary typicality used there as the likelihood of the perception $p$
given the hypothesis of a particular value of $n$, lead one to a very small
posterior probability that $n\gg 1$ even if one started with the unnormalized
uniform prior distribution $P(n)dn = dn$ that is almost entirely weighted at
arbitrarily large values of $n$.

 If we average the posterior probability distribution Eq.~(\ref{eq:14i}) over
the gaussian distribution $e^{-p^2/2}$ that would be given for $p$ if indeed
SQM and its value of $n=1$ were correct, then one would get the following
averaged posterior distribution for $n$:
 \begin{equation}
 \bar{P}(n)dn = 2(\arctan{{1 \over \sqrt{n}}}
 -{\sqrt{n}\over n+1})dn.
 \label{eq:14jj}
 \end{equation}
This distribution is normalized, unlike the prior distribution for $n$ that was
adopted, but since for large $n$, $\bar{P}(n)\sim (4/3)n^{-3/2}$, the mean and
higher moments for $n$ are infinite.  Therefore, if all equal ranges for $n$
are {\it a priori} assumed to be equally likely, then the average of all
observations of $p$, if $n=1$ were really unknowingly correct, would damp the
posterior probability distribution for $n$ at large $n$ so that it would become
normalizable, but it would be damped so weakly that the mean would still be
undefined.  Nevertheless, any single observation of a $p\neq 0$ would give a
posterior distribution Eq.~(\ref{eq:14i}) exponentially damped for sufficiently
large $n$, and hence with a finite mean and rms value, even though these finite
values from individual observations would average out to divergent values when
averaged over the distribution of $p$ that would result if indeed $n=1$ (or
indeed if $n$ were any other precise positive value).  In any case, with or
without the averaging over the values of $p$, the posterior probability would
be very small that $n$ would have a value that is sufficiently large, giving
evidence against such large values of $n$.

 On the other hand, the posterior probability distribution $P(n|p)dn$ of
Eq.~(\ref{eq:14i}) does not provide significant evidence against a value of $n$
much smaller than $1/p^2$, since it is even relatively larger at very small
values of $n$ than was the uniform $P(n)dn$ from which it was derived.  This
illustrates the limitations of using merely the ordinary typicality to deduce
posterior probabilities, since it provides no penalty for results ``too good to
be true,'' namely ordinary typicalities very near unity.  In this example, if
$n$ were very small, a $p$ near one would be much closer to its mean of zero
than the standard deviation $1/\sqrt{n}$ for $p$ in the distribution
$f[m(p)]=m(p)^n=e^{-np^2/2}$.  Intuitively, we ought be be surprised if we get
a result much closer to the peak of a gaussian probability distribution than
one standard deviation, but using only the ordinary typicality does not capture
this intuition, since it says we should only be surprised if we get a result
too many standard deviations from the mean.

 Using purely the reversed typicality $T_r(p)$ instead would not be any good
here, since it would not give any penalty for getting a result with very low
ordinary typicality.  In fact, the denominator of Eq.~(\ref{eq:15}) would then
simply diverge from the integration over negative $n$, which would give a
reversed typicality of unity.  If one put a cutoff at large negative $n$, did
the calculation, and then let the cutoff tend to negative infinity, one would
find that the resulting $P_r(n|p)$ would have almost all its contribution from
arbitrarily negative $n$.

 The best of the three typicalities to use in Eq.~(\ref{eq:15}) would thus
appear to be the dual typicality $T_d(p)$.  Inserting this and the uniform
prior distribution for $n$ into Eq.~(\ref{eq:15}) gives
 \begin{equation}
 P_d(n|p)dn = Np^2
 \min{[{\rm erfc}(\sqrt{p^2 n/2}),
 {\rm erf}(\sqrt{p^2 n/2})]}dn,
 \label{eq:14k}
 \end{equation}
where
 \begin{equation}
 N^{-1} = 1 - 2x_1^2 + 8\int_{0}^{x_1}{dx\,x\,{\rm erfc}(x)}
 \approx 0.857348\approx (1.166387)^{-1}.
 \label{eq:14m}
 \end{equation}
with $x_1\approx 0.476936$ being the positive value of $x$ for which ${\rm
erf}(x)={\rm erfc}(x)=1/2$.  This posterior probability distribution
$P_d(n|p)dn$ is not only damped at large $p^2 n$, which would give $p$ a low
ordinary typicality, but also at small $p^2 n$, which would give $p$ a low
reversed typicality (i.e., an ordinary typicality unusually near one).
$P_d(n|p)dn$ gives a mean value for $n$ of about $1.727468/p^2$, and a standard
deviation of about $1.686141/p^2$, which are both even higher than those for
$P(n|p)dn$, essentially because $P_d(n|p)dn$ damps the contribution at small
$n$ and gives a higher weight to the contribution for larger $n$.

 Thus we may note that the damping at small $p^2 n$ is so weak, going only as
the square root of $n$, that a perception $p$, even if it is reasonably close
to one standard deviation from the mean, does not put very tight limits on $n$
in SQM$n$ with a uniform prior distribution for $n$.  In this way it seems, at
least for a gaussian distribution of expectation values for experience or
perception operators $E(p)$ (or for a discrete distribution that is
approximately gaussian, such as a binomial distribution for a large number of
possibilities), that if one allows the measures for the perceptions to be an
arbitrary power $n$ of the expectation values with a broad prior distribution
for $n$, then no observation (i.e., perception) can give very tight limits on
$n$.  Thus it may be that there is actually very little evidence (except for
the simplicity that leads me to propose $n=1$ in Sensible Quantum Mechanics
itself) that probabilities in quantum mechanics (which I have argued apply only
to conscious perceptions) are proportional to the {\it squares} of the absolute
values of appropriate amplitudes (i.e., to the {\it first} power of the
expectation values of positive perception operators).  It would be interesting
to see whether there are any highly nongaussian distributions for perceptions
that would be suitable for putting stringent limits on the power $n$ of
the expectation value that enters in SQM$n$ theories.

 One idea for testing the exponent $n$ more stringently
in SQM$n$ theories is the following:
Consider a set of perceptions that may be divided up into
a large number $N$ of subsets, and for which one has some
control on the relative expectation values of the experience operators $E(p)$.
For example, consider the set $S$ of perceptions that include a conscious
awareness of all the digits of a nonnegative decimal integer
with no more than $k$ digits.
(It seems hard for me, by looking at numbers on a computer terminal,
to convince myself that I can be consciously aware, in a single
simultaneous perception, of the values of more than about $k=8$ digits,
and I am not sure I can really be conscious of all those at once,
but $k=8$ is my rough subjective estimate of the limit for me.
Note that this is {\it not} the number of digits I can memorize,
for I need not have all the digits that I can remember
ever simultaneously be in any single conscious perception.)

 This set of perceptions can now be divided up into the $N=10^k$
subsets of perceptions that each include a component of being
consciously aware of the values of the decimal digits of a particular
nonnegative integer of $k$ digits or less.
Group these $N=10^k$ sets into three sets,
the first ($S_1$) containing the perceptions of a particular subset of $N_1$
of the integers, the second ($S_2$) containing the perceptions of a second
subset of $N_2$ integers (not overlapping with the first),
and the third ($S_3$) containing the the perceptions of the
remaining $N_3 = N - N_1 - N_2$ integers.

 Now employ some quantum decision-making device (such as
a nonalgorithmic random number generator that invokes
quantum measurements and is nondeterministic in the usual sense,
even if it is deterministic in my global view when one considers its
effects within the entire wavefunction, i.e., across all the Everett
many worlds in that crude way of describing things, but not within
a single randomly chosen Everett world, where it will appear random).
Use this device to produce
a quantum expectation value for each of the integers to appear, by itself,
on a computer terminal or printout where it can be read,
with the expectation value being roughly independent of the
particular integer within each of the three sets described above,
but depending on which set the integer is in.

 There is then the process of reading the integer and transferring
the information to whatever part of the brain that produces
the conscious experience that includes the awareness of
all the digits of the integer (somewhat more accurately,
the part of the brain that has the relevant structure
for the experience operator $E(p)$ for each of the possible perceptions
of the entire integer).  If this process can be idealized as making
$m(p) = \langle E(p) \rangle \equiv \sigma[E(p)]$ proportional to
the quantum expectation value for the corresponding integer
to appear on a terminal or printout
(an expectation value that, up to a constant of proportionality,
can be controlled by the experimenter),
then one can choose the relative values of the $m(p)$'s,
say $m_1$, $m_2$, and $m_3$ respectively,
for each of the three sets of integers.

 Now in ordinary Sensible Quantum Mechanics, the measure density
for each perception $p$ is simply the corresponding $m(p)$,
but in the SQM$n$ alteration of SQM given by Eq.~(\ref{eq:12b}),
the measure density is $f[m(p)]=m(p)^n$, where the exponent $n$
may be different from the value of 1 that it is postulated to have
in pure SQM.  Therefore, in SQM$n$, the conditional probability
that a perception is in the set $S_2$, say, given that it is in the
set $S = S_1\cup S_2\cup S_3$, is, under the idealizations above,
 \begin{equation}
 P_n(S_2|S)\equiv {\mu(S_2) \over \mu(S)}
 = {m_2^n N_2 \over m_1^n N_1 + m_2^n N_2 + m_3^n N_3},
 \label{eq:14n}
 \end{equation}
which depends on $n$ and the ratios of the $m$'s,
though not on the overall normalization of the $m$'s,
which is arbitrary.

 For example, let $N_1 = 1$, $N_2 = 10^{k/2}$, so that then
(assuming $10^k \gg 1$, as it is for, say, $k = 8$)
$N_3 = 10^k - 10^{k/2} - 1 \approx 10^k = N$.
Then if one selects the quantum decision-making device
so that, up to an arbitrary constant overall normalization factor, 
the $m$'s are
$m_1 = 1/N_1 = 1$, $m_2 = 1/N_2 = 10^{-k/2}$, and
$m_3 = 1/N_3 \approx 10^{-k}$, one gets
 \begin{equation}
 P_n(S_2|S) \approx {1 \over 10^{(n-1)k/2} + 1 + 10^{-(n-1)k/2}}.
 \label{eq:14o}
 \end{equation}

 One can now see from this that if one has ordinary SQM with $n=1$,
there is roughly one chance in three that a perception in the set $S$
will be in the particular subset $S_2$, so such a perception would not
be atypical if it occurred.  But if such a perception occurred
and one assumed that an SQM$n$ theory applied with $n$ significantly
different from 1, then the probability that a perception within $S$
is also within $S_2$ is only roughly $10^{-|n-1|k/2}$, e.g.,
only $10^{-4}$ if $k=8$ and either $n=0$ or $n=2$.  Therefore, such
a perception would be highly atypical within such a theory,
and one would then have strong statistical evidence for rejecting
such a theory, unless one had assigned it a prior probability $P(H_i)$
that were very much nearer unity than the prior probability assigned
to ordinary SQM with $n=1$.
In another way of stating it, at the 99\% confidence level
(i.e., rejecting hypotheses that predict that the conditional probability
of the observed perception within $S_2$ is less than 1\%),
the value of the exponent $n$ would obey $|n-1|
{\ \lower-1.2pt\vbox{\hbox{\rlap{$<$}\lower5pt\vbox{\hbox{$\sim$}}}}\ }
4/k$.

 It is somewhat discouraging that this bound, if it indeed can be found
to be true, is not very tight, say if $k$ is roughly 8, but at least it is
sufficient to rule out at the 99\% confidence level what might be seen as
the simplest alternative to SQM with $n=1$, namely SQM$n$ with $n=2$.
(The theory with $n=0$ could probably be ruled out by other considerations,
such as not predicting that it is any more typical to perceive getting
a million heads than to get zero after throwing a fair coin two million times,
if one is not perceiving the order of the million heads, and if one can avoid
the Attention Effect that would no doubt amplify the measure
for perceptions of zero heads.)

 It may also be somewhat discouraging to note that with the
quantum expectation values, the $m$'s, chosen as above,
there is a two-thirds conditional probability even within SQM
that the perception will not be within the set $S_2$.
If so, this test would tell one virtually nothing, since then the perception
(or at least the fact that it is in $S$ but not in $S_2$)
would be typical for any SQM$n$ theory.
One might want to select the quantum decision-making device
to make $m_2$ relatively larger than the choice above,
so that then if SQM is correct the perception within $S$
will almost certainly also be within $S_2$, but then the
conditional probability that it would be within $S_2$
even if SQM$n$ were true with a different value of $n$
would also be higher, so the test would be less
sensitive in ruling out different values of $n$.

 Of course, if one wanted instead to try to show that $n$
is {\it not} equal to 1, one might make $m_2$ sufficiently large
that a perception within $S$ would almost certainly
also be within $S_2$ if $n=1$, and then if it were found
that such a perception were not within $S_2$,
this would be statistical evidence that $n$ is not one.
Given the high prior probability that one might tend to assign
to $n=1$, it might take a perception outside $S_2$ with $m_2$
very high in order to make much of a case for concluding that
actually $n$ is not equal to unity.

 It is important to note that, given the irreducibly first-person
nature of conscious experience, it would not be sufficient
evidence for anyone to be told of the result of such an experiment
to test the value of $n$,
since in SQM$n$ one's consciousness of being told of the result
would have a measure that is proportional to the $n$th power of
the expectation value of the experience operator $E(p)$
for the consciousness of being told, not of the $E(p)$
for the consciousness of the digits of the decimal integer.
Anyone who wished to experience the result of the experiment
would need to be directly conscious of the digits themselves
(which could of course be printed in the report of the experiment,
so that each reader could in a sense do the consciousness part
of the experiment for himself or herself).
It would also not even be sufficient for an experimenter
simply to {\it remember} having done the experiment and
having, say, gotten a result supporting $n$ equal to or close to 1,
if he or she were not in the same perception consciously aware of
the digits of the number itself, because what would matter for that
conscious memory would be the $n$th power of
the expectation value of the experience operator $E(p)$
for the consciousness of the memory, not of the $E(p)$
for the consciousness of the digits of the decimal integer itself.
This is another aspect that makes it apparently very difficult
to get much strong evidence about the value of the exponent $n$
in the generalization of Sensible Quantum Mechanics to SQM$n$
theories.

\section{Toy Models for Sensible Quantum Mechanics}

\hspace{.25in}To illustrate some of the structures and hypotheses discussed
above for SQM itself, let us first consider a simple toy system for which the
quantum world has a Hilbert-space dimension of two, i.e., the spin states of a
spin-half system with basis states $|1\rangle = |\sigma_z=1\rangle =
|\!\uparrow\rangle$ and $|2\rangle = |\sigma_z=-1\rangle =
|\!\downarrow\rangle$.  A general positive operator for this system has the
matrix form
 \begin{equation}
 \pmatrix{t+z & x+iy \cr x-iy & t-z \cr}
 \label{eq:18}
 \end{equation}
for
 \begin{equation}
 t\geq \sqrt{x^2 + y^2 + z^2}.
 \label{eq:19}
 \end{equation}
Hence any awareness operator $A(S)$ for any set of perceptions must have this
form for this two-state quantum system.  The set of parameters $\{t,x,y,z\}$
with the inequality (\ref{eq:19}) forms a four-dimensional manifold with
boundary, which one can readily visualize in this example as the interior of
the future light cone, and its boundary, in a (fictitious) four-dimensional
Minkowski spacetime.  For the $A(S)$ to be a positive-operator-valued measure
as the Quantum-Consciousness Connection assumption demands, the parameters
$\{t,x,y,z\}$ must be linear functions of the sets $S$, so that the parameters
for the union of disjoint subsets are the sums of the corresponding parameters
for the subsets themselves.

 First let us consider SQMC theories in which the set of perceptions is
continuous.  If this set forms a manifold, one can assign a volume element
$d\mu_0(p)$ to it to give the prior measure (\ref{eq:4}), from which by
Eq.~(\ref{eq:5}) one can define an experience operator $E(p)$ for each
perception $p$, also necessarily of the form (\ref{eq:18}) as each $A(S)$ is.

 If the dimensionality of the set $M$ of perceptions in this model is larger
than three, and if the map from this set to the set of (necessarily positive)
experience operators $E(p)$ is smooth, then necessarily two different
perceptions must have experience operators proportional to each other.  Thus
the Pairwise Independence Hypothesis will not hold in this case.

 In order to require that the Pairwise Independence Hypothesis hold, I shall
henceforth assume that the dimensionality of the set $M$ of perceptions in this
toy model is no larger than three (i.e., one less than the square of the
dimension of the quantum Hilbert space being considered).  First, consider the
case in which the dimensionality is exactly three, and for simplicity assume
that perceptions in the set are parametrized by the triplet $\{u,v,w\}$ of real
numbers obeying the inequality
 \begin{equation}
 r(u,v,w) \equiv \sqrt{u^2 + v^2 + w^2}\leq 1.
 \label{eq:20}
 \end{equation}
Then one simple choice for the experience operators is
 \begin{equation}
 E(p)\equiv E(u,v,w)=\pmatrix{t(1+w) & t(u+iv) \cr t(u-iv)  & t(1-w) \cr},
 \label{eq:21}
 \end{equation}
where $t=t(u,v,w)$ is some nonnegative weight function over the set of
perceptions.  One can also write the prior measure volume element as
 \begin{equation}
 d\mu_0(p) = m_0(u,v,w) du dv dw
 \label{eq:22}
 \end{equation}
for some nonnegative weight function $m_0(u,v,w)$.  Then Eq.~(\ref{eq:5}) gives
 \begin{equation}
 A(S) = \int_S t(u,v,w) m_0(u,v,w) du dv dw
 \pmatrix{1+w & u+iv \cr u-iv & 1-w \cr}.
 \label{eq:23}
 \end{equation}

 One sees that the awareness operators do not depend separately upon the weight
functions $t(u,v,w)$ and $m_0(u,v,w)$, but only upon their product.  Of course,
the prior measure
 \begin{equation}
 \mu_0(S) \equiv \int_{S}{d\mu_0(p)}
 = \int_S m_0(u,v,w) du dv dw
 \label{eq:24}
 \end{equation}
does depend upon $m_0(u,v,w)$ alone, so it is logically an independent degree
of freedom.  However, in this model, one could adopt one of the normalization
hypotheses for the experience operators to fix $t(u,v,w)$.  For example, the
Constant-Maximum-Normalization Hypothesis leads to $t=1/[1+r(u,v,w)]$; the
Unit-Normalization Hypothesis, which is equivalent to the hypothesis
(\ref{eq:9}) of SQMCT, leads to $t(u,v,w)=1/2$; and the
Projection-Normalization Hypothesis leads to
$t=1/[1+r(u,v,w)^2]=1/(1+u^2+v^2+w^2)$.  Furthermore, one could use
Eqs.~(\ref{eq:11}) and (\ref{eq:12}) of SQMCR to get $m_0(u,v,w)$ in terms of
$t(u,v,w)$ and its first derivatives (in a slightly complicated formula not
worth copying here for the general $t(u,v,w)$).  For example, combining
Eqs.~(\ref{eq:11}) and (\ref{eq:12}) with the Projection-Normalization
Hypothesis leads to $m_0(u,v,w)=\sqrt{8}/(1+u^2+v^2+w^2)^3$, which is what I
shall take for concreteness in the following discussion of this example.

 Now if one takes the quantum state to have the pure state
$\rho=|1\rangle\langle 1|$, then $m(p)=\langle E(p) \rangle = t(1+w) =
(1+w)/(1+u^2+v^2+w^2)$.  One can insert this into Eq.~(\ref{eq:6}) to get the
measure for any set of perceptions, with the set $S$ here being given by some
region of the three-dimensional space with coordinates $(u,v,w)$:
 \begin{equation}
 \mu(S) = \int_S m(u,v,w) m_0(u,v,w) du dv dw
 = \int_S {\sqrt{8}(1+w) \over (1+u^2+v^2+w^2)^4}
 du dv dw.
 \label{eq:25}
 \end{equation}
One could then take appropriate ratios of such measures, such as in
Eq.~(\ref{eq:3}), as giving conditional probabilities for various sets of
perceptions.  For example, one could calculate the probability (\ref{eq:14})
that the typicality of a random perception in this measure is less than that of
a particular perception $p$ labeled by $(u,v,w)$.  For a generic perception in
this present example, this calculation appears to be too messy to be worth
doing here, but the point is that, given the assumptions made above, it can in
principle be done (perhaps numerically if it is not possible to give explicit
elementary formulas for the result).

 Another example within SQMC that one might consider is the case in which the
set $M$ of perceptions forms a manifold of dimension one, say the circle $S^1$
parametrized by the angle $\phi$ that runs from zero to $2\pi$ and then
repeats.   (One might regard this angle as denoting a cyclic time as perceived,
but of course this would depend on what is the precise content of the
perception $p$ parametrized by the angle $\phi$.  The contents of the
perceptions are features that are not captured merely by the topology of the
set $M$, the prior measure, and the experience operators.)  In this case one
can adopt the Projection Hypothesis as well as the Pairwise Independence
Hypothesis, giving SQMP (or, to state more explicitly the assumption of the
Pairwise Independence Hypothesis, SQMIP).  For example, the experience
operators can take the form Eq.~(\ref{eq:21}) with $t=1/2$, $u=\cos{\phi}$,
$v=\sin{\phi}$, and $w=0$, giving the projection operators
 \begin{equation}
 E(p)\equiv E(\phi)=P(\phi)=
 {1 \over 2}\pmatrix{1 & e^{i\phi}\cr e^{-i\phi}& 1 \cr}.
 \label{eq:26}
 \end{equation}
If we write the prior measure volume element as
 \begin{equation}
 d\mu_0(p) = m_0(\phi) d\phi,
 \label{eq:27}
 \end{equation}
for some nonnegative weight function $m_0(\phi)$, then Eq.~(\ref{eq:5}) gives
 \begin{equation}
 A(S) = \int_S m_0(\phi) d\phi
 {1 \over 2}\pmatrix{1 & e^{i\phi}\cr e^{-i\phi}& 1 \cr}.
 \label{eq:28}
 \end{equation}
In this case of a one-dimensional set of perceptions and the experience
projection operators of Eq.~(\ref{eq:26}), the only natural choice for the
prior measure density $m_0(\phi)$ is a constant, which is what will be assumed
here.  For example, Eqs.~(\ref{eq:11}) and (\ref{eq:12}) of SQMCR give
$m_0(\phi)=1/\sqrt{2}$, but the value of the constant is not relevant to
testable conditional probabilities or typicalities.

 If the quantum state is the pure state \\$\rho=(\cos{{1 \over
2}\theta}|1\rangle + \sin{{1 \over 2}\theta}|2\rangle) (\cos{{1 \over
2}\theta}\langle 1| + \sin{{1 \over 2}\theta}\langle 2|)$, giving the density
matrix
 \begin{equation}
 \rho=
 {1 \over 2}\pmatrix{1+\cos{\theta} & \sin{\theta}\cr  \sin{\theta} &
1-\cos{\theta} \cr},
 \label{eq:29}
 \end{equation}
then
 \begin{equation}
 m(p)=\langle E(p) \rangle=Tr(E(p)\rho)
 ={1 \over 2}(1+\sin{\theta}\cos{\phi}).
 \label{eq:30}
 \end{equation}
Assuming for simplicity that $\sin{\theta}>0$, this leads to the following
expressions for the ordinary, reversed, and dual typicalities:
 \begin{eqnarray}
 T(p) & = & 1-|\phi+\sin{\theta}\sin{\phi}|/\pi,\\
 T_r(p) & = & |\phi+\sin{\theta}\sin{\phi}|/\pi,\\
 T(p) & = & 1-|1-2(|\phi+\sin{\theta}\sin{\phi}|/\pi)|.
 \end{eqnarray}
For example, if it were hypothesized that $\theta=\pi/2=90^{\circ}$, and one
observed $\phi=5\pi/6=150^{\circ}$, then $T(p)=(\pi-3)/(6\pi)\approx 0.007512$,
so by the criterion that the typicality should not be too small, one could rule
out this hypothesis for $\theta$ at the 99\% confidence level.

 Although we have been using the Heisenberg picture, in which the state is
fixed, it might be helpful to think of the present example in the
Schr\"{o}dinger picture in which the experience operator $E(p)$ is held fixed
as a function of the ``perceived time'' $\phi$, and instead the state $\rho$
changes with this ``time.''  In this example, one can then think of the state
as representing the direction of the spin of a spin-half particle which
precesses around the equatorial plane perpendicular to an axis at an angle
$\theta$ from the direction corresponding to the experience operator.  When the
spin direction is closer to that of the experience operator, one gets a larger
measure density $m(p)$ for the corresponding perception.  This example
illustrates how the measure density need not be constant as a function of the
``perceived time,'' as Sensible Quantum Mechanics has no requirement of any
``unitarity'' in the sense of conservation with ``time'' of any probability or
measure or measure density.

 One could still maintain the Projection Hypothesis and yet extend the example
to one in which the perceptions form a two-dimensional space.  For example, the
experience operators could take the form Eq.~(\ref{eq:21}) with $t=1/2$,
$u=\sin{\vartheta}\cos{\varphi}$, $v=\sin{\vartheta}\sin{\varphi}$, and
$w=\cos{\vartheta}$, giving the projection operators
 \begin{equation}
 E(p)\equiv E(\vartheta,\varphi)=P(\vartheta,\varphi)=
 {1 \over 2}\pmatrix{1+\cos{\vartheta} &
 \sin{\vartheta}e^{i\varphi}\cr
 \sin{\vartheta}e^{-i\varphi}& 1-\cos{\vartheta} \cr}.
 \label{eq:31}
 \end{equation}
Here $(\vartheta,\varphi)$ are polar coordinates for a two-sphere.  If one
restricts $\vartheta$ not to include the values of 0 and $\pi$ that it would
take at the poles of the sphere, one could take $\varphi$ to be a ``perceived
time,'' as $\phi$ was in the previous example. Then one can take $\vartheta$ to
be an a second component of the perception, e.g., a ``perceived temperature.''
One might even divide up the space of perceptions so that those with
$0<\vartheta<\pi/2$ are defined to be those of a ``cold'' individual mind, and
those with $\pi/2<\vartheta<\pi$ are defined to be those of a ``hot''
individual.  (The ``lukewarm'' mind with $\vartheta=\pi/2$ forms a set of
perceptions of measure zero and will be henceforth ignored.)  However, as
mentioned above, this division of perceptions into individual ``minds'' is {\it
ad hoc} and not fundamental to Sensible Quantum Mechanics.

 If one takes the rotationally-invariant prior measure volume element as
 \begin{equation}
 d\mu_0(p) = \sin{\vartheta} d\vartheta d\varphi,
 \label{eq:32}
 \end{equation}
which is proportional to what Eqs.~(\ref{eq:11}) and (\ref{eq:12}) of SQMCR
would give, then Eq.~(\ref{eq:5}) gives
 \begin{equation}
 A(S) = \int_S \sin{\vartheta} d\vartheta d\varphi
 {1 \over 2}\pmatrix{1+\cos{\vartheta} &
 \sin{\vartheta}e^{i\varphi}\cr
 \sin{\vartheta}e^{-i\varphi}& 1-\cos{\vartheta} \cr}.
 \label{eq:33}
 \end{equation}

 Again taking the quantum state to be given by the (pure) density matrix of
Eq.~(\ref{eq:29}), the measure density for each perception is
 \begin{eqnarray}
 m(p) \equiv m(\vartheta,\varphi)
 =\langle E(p) \rangle=Tr(E(p)\rho)
 \nonumber \\
 = {1 \over 2}(1+\cos{\theta}\cos{\vartheta}
 +\sin{\theta}\sin{\vartheta}\cos{\varphi})
 =\cos^2{{1 \over 2}\psi},
 \label{eq:34}
 \end{eqnarray}
where $\psi$ is the angle between the spin direction corresponding to the state
$\rho$, at polar coordinates $(\theta, \phi=0)$, and that corresponding to the
experience operator $E(p)$, at polar coordinates $(\vartheta, \varphi)$.  Then
the measure for any set $S$ of perceptions, here being given by some region of
the two-sphere, is
 \begin{eqnarray}
 \mu(S) = \int_S m(p) d\mu_0(p)
 = \int_S m(\vartheta,\varphi)
 \sin{\vartheta} d\vartheta d\varphi
 \nonumber \\
 = \int_S {1 \over 2}(1+\cos{\theta}\cos{\vartheta}
 +\sin{\theta}\sin{\vartheta}\cos{\varphi})
 \sin{\vartheta} d\vartheta d\varphi.
 \label{eq:35}
 \end{eqnarray}

 From this measure, one can readily calculate the various typicalities of each
perception as
 \begin{eqnarray}
 T(p) & = & \cos^4{{1 \over 2}\psi},\\
 T_r(p) & = & 1-\cos^4{{1 \over 2}\psi},\\
 T(p) & = & 1-|1-2\cos^4{{1 \over 2}\psi}|.
 \end{eqnarray}
One can also calculate various conditional probabilities, such as the
conditional probability distribution for the ``perceived temperature''
$\vartheta$ at a fixed range (perhaps infinitesimal) of the ``perceived time''
$\varphi$, or even the inverse conditional probability distribution for the
``time'' at a fixed ``temperature.''  (One easily sees there need be no
preference for the perceived time component of a perception to be used as a
condition, rather than for any other component or aspect of a perception to be
thus used.  That is, Sensible Quantum Mechanics not only does not need a
preferred time variable in the description of the quantum state and operators,
but also it does not need to single out a preferred temporal aspect of
perceptions on which to base all conditional probabilities.)

 There are also other probabilities that one can calculate, such as the
probability that a perception belongs to a ``cold'' individual, which, using
the {\it ad hoc} division of the perceptions into those of ``cold'' and ``hot''
individuals above, comes out to be $(2+\cos{\theta})/4$.  In a similar way, in
a complete Sensible Quantum Mechanics theory, and with a precisely-defined {\it
ad hoc} division of the possible perceptions into those of, say, humans, dogs,
insects, electronic computers, etc., one (with this one admittedly being
probably only a sufficiently intelligent being outside our universe) should in
principle be able to calculate the relative (frequency-type) probabilities that
a random perception fits into one of these categories.  It would certainly be
interesting to know what these relative probabilities are (for some reasonable
choice of the classification).  Since my own perception is human (or at least I
presently perceive it to be), and since I see no reason why it should not be
typical, I personally would suspect (even though one can readily see that this
suspicion does not directly follow merely from the assumption that the
typicality of my perception is not small) that the probability of a human
perception would be greater than, or at least of a comparable magnitude to,
that of dogs, insects, or electronic computers, even including those with
perceptions that they are far to the future of what we perceive the present
epoch to be.

 Next, let us consider SQMD theories in which the set of perceptions is
discrete, but still for the moment continue to use the simple toy system for
which the quantum world has a Hilbert-space dimension of two.  If we adopt the
Pairwise Independence Hypothesis and the Projection Hypothesis, thus getting
SMQIP theories, but do not also adopt the Linear Independence Hypothesis to get
the Linearly Independent Projection Hypothesis and the resulting SQMLIP
theories, then the experience operators $E(p)$ can be projection operators of
the form given by Eq.~(\ref{eq:31}) corresponding to each of any discrete set
of directions in three-space or points on a two-sphere (e.g., points each
labeled by a pair of polar coordinates $(\vartheta,\varphi)$).   (One could
also have one of the experience operators being equal to the identity operator,
 the only rank-two projection operator for the present toy system.)  There can
be an arbitrary large number (even a countably infinite number, or even an
uncountably infinite number) of such pairwise independent discrete experience
projection operators, so without further hypotheses they are not limited by the
dimension of the quantum Hilbert space.  As a result, there can be an
arbitrarily large number of measure ``densities'' $m(p)$ even in the discrete
case.

 If the set of these distinct experience projection operators is not fixed,
then for any pure quantum state such as the one given by the density matrix of
Eq.~(\ref{eq:21}), the ratios of all the $m(p)$'s can be arbitrary.  (If the
state is not pure, then the ratios are bounded by the ratio of the larger to
the smaller eigenvalue of the density matrix.)  On the other hand, if the set
of $E(p)$'s {\it is} fixed, and there is only the freedom of what the state is
(including how impure it may be), then since there are at most four linearly
independent positive operators (or projection operators) in the Hilbert space
of dimension two, there are at most four independent $m(p)$'s and three
independent ratios of them.  In general, even if the quantum world is described
by a Hilbert space of finite dimension $N$, with $N\times N$ positive,
hermitian density matrices as states, and even if there is an arbitrarily
large but discrete number of possible perceptions $p$ and corresponding
experience operators $E(p)$, the corresponding measures $m(p)$ for these
perceptions can have arbitrary ratios for {\it any} pure state and hence do
{\it not} determine the state if the experience operators themselves are not
known.  On the other hand, if the experience operators are known, then there
are at most $N^2$ linearly independent $m(p)$'s (or $N^2-1$ if the density
matrix is assumed to be normalized), so for a generic set of known $E(p)$'s,
$N^2-1$ ratios of $m(p)$'s would uniquely determine the state up to
normalization.

 Of course, in reality, any conscious being within the system has access to
only one perception and does not even have access to its measure (except that
he may typically assume that the measure is high enough that it is not too
atypical), so he can only use something like the Bayesian reasoning of
Eq.~(\ref{eq:15}) with some assignment of prior probabilities (say based on
simplicity) in order to make an effectively probabilistic guess of the correct
theory (e.g., of the correct experience operators and the correct quantum
state, if the framework of Sensible Quantum Mechanics is assumed).

 (Perhaps it would be more nearly true to say that if SQM is true, the correct
experience operators and the correct quantum state simply directly determine
the perceptions, and their measures, of beliefs in various theories, but in my
perception they seem to be determining that I come up with some sort of
rational idealization of how this choice might be made by an agent that is
truly free to assign probabilities to theories based on Bayesian reasoning.  It
is a deeper mystery how such a free agent may be an idealized approximation to
us when we choose theories if our thought processes are not really free to
choose to follow logical reasoning, say starting with certain Bayesian
assumptions, including a set of prior probabilities for various hypotheses.)

 Because there are only $N^2-1$ independent ratios of $m(p)$'s for a fixed set
of experience operators in a Hilbert space dimension of dimension $N$, one
might prefer (though I myself do not see a strong reason for this preference)
to restrict the number of experience operators to be no greater than $N^2$ and
to require that they obey the Linear Independence Hypothesis, which, once the
Projection Hypothesis is also assumed, leads to SQMLIP theories obeying the
Linearly Independent Projection Hypothesis.  For the simple two-dimensional
Hilbert-space model we have been considering, one can have four rank-one
projection operators of the form given by Eq.~(\ref{eq:31}), with the
corresponding four directions in three-space not all being coplanar, or,
equivalently, with the corresponding points on the two-sphere not all being on
the same great circle.  If one has instead the identity operator and three
rank-one projection operators, then the corresponding three directions in
three-space must not be coplanar, or the corresponding points on the two-sphere
must not all be on the same great circle.

 If one makes the Commuting Projection Hypothesis instead of the Linearly
Independent Projection Hypothesis, then for the two-dimensional Hilbert-space
model one can have at most three experience operators, the identity operator
and two orthogonal rank-one projection operators (i.e., two corresponding to
two opposite directions in three-space, or to antipodal points on the
two-sphere).  These three are of course not linearly independent, since the two
orthogonal rank-one projection operators add up to the identity operators, so
this particular SQMPC theory does not obey the Linearly Independent Projection
Hypothesis.  If one added the latter, one could have at most two experience
operators, either the identity operator and one (arbitrary) rank-one projection
operator, or else two orthogonal rank-one projection operators.

 As a final example obeying the Projection Hypothesis, one may adopt the
Orthogonal Projection Hypothesis, which is the strongest of the possible
hypotheses listed above for the experience operators and which implies all the
others (except the Unit-Normalization Hypothesis, which is inconsistent with
the Projection Hypothesis unless all the projection operators are of rank one).
 Then one could either have up to two orthogonal rank-one projection operators,
or the one identity operator, for the set of experience operators.  Only in the
former case does the Unit-Normalization Hypothesis also hold.

 If in the discrete SQMD case we weaken the Projection Hypothesis to the
Sequence of Projections Hypothesis, then without further restrictions we can
have an arbitrarily large number of experience operators, even if we continue
to assume the Pairwise Independence Hypothesis.  Of course, if we also make the
Linear Independence Hypothesis, then we are again limited to at most $N^2$
experience operators (e.g., four for the two-dimensional Hilbert-space model),
but in the following I shall not assume this.

 Suppose we consider the case in which each sequence has two rank-one
projection operators, of which the first can be, in matrix notation, either
 \begin{equation}
 Q={1 \over 2}\pmatrix{1+\cos{\theta_1} &
 \sin{\theta_1}e^{i\phi_1}\cr
 \sin{\theta_1}e^{-i\phi_1}& 1-\cos{\theta_1} \cr}
 \label{eq:36}
 \end{equation}
or $I-Q$, and of which the second can be either
 \begin{equation}
 R={1 \over 2}\pmatrix{1+\cos{\theta_2} &
 \sin{\theta_2}e^{i\phi_2}\cr
 \sin{\theta_2}e^{-i\phi_2}& 1-\cos{\theta_2} \cr}
 \label{eq:37}
 \end{equation}
or $I-R$.  The resulting set of four sequences are thus
 \begin{eqnarray}
 C(1) = & RQ,\\
 C(2) = & R(I-Q) & = R-RQ,\\
 C(3) = & (I-R)Q & = Q-RQ.\\
 C(4) = & (I-R)(I-Q) & = I-Q-R+RQ,
 \end{eqnarray}
and by the Sequence of Projections Hypothesis the experience operators are then
 \begin{eqnarray}
 E(1) = & C(1)^{\dagger}C(1) & = QRQ,\\
 E(2) = & (I-Q)R(I-Q) & = R-QR-RQ+QRQ,\\
 E(3) = & Q(I-R)Q & = Q-QRQ,\\
 E(4) = & (I-Q)(I-R)(I-Q) & = I-Q-R+QR+RQ-QRQ.
 \end{eqnarray}

 Now assume that the state is given by the pure-state density matrix (i.e.,
another rank-one projection operator)
 \begin{equation}
 \rho={1 \over 2}\pmatrix{1+\cos{\theta_0} &
 \sin{\theta_0}e^{i\phi_0}\cr
 \sin{\theta_0}e^{-i\phi_0}& 1-\cos{\theta_0} \cr}.
 \label{eq:38}
 \end{equation}
Then in this case the Consistent Sequence of Projections Hypothesis, the
Individually Weak Decoherent Histories Hypothesis, and the Weak Decoherent
Histories Hypothesis all give the same single real equation
 \begin{equation}
 \sigma[QR+RQ-2QRQ]\equiv
 2Re\,Tr[(QR-QRQ)\rho] = 0.
 \label{eq:39}
 \end{equation}
Similarly, the Individually Medium Decoherent Histories Hypothesis, the Medium
Decoherent Histories Hypothesis, and the Individually Strong Decoherent
Histories Hypothesis all give the one complex equation
 \begin{equation}
 \sigma[QR-QRQ]\equiv
 Tr[(QR-QRQ)\rho] = 0.
 \label{eq:40}
 \end{equation}
In the present example, the Strong Decoherent Histories Hypothesis is
impossible to satisfy, since it would require four orthogonal projection
operators (one for each sequence) in the two-dimensional Hilbert space.
Finally, the Linearly Positive Histories Hypothesis gives the inequality
 \begin{equation}
 \max{(0,\sigma[Q+R-I])}\leq
 Re\,\sigma[QR]\equiv
 Re\,Tr(QR\rho)\leq
 \min{(\sigma[Q],\sigma[R])}.
 \label{eq:41}
 \end{equation}

 One can express these conditions (\ref{eq:39})-(\ref{eq:41}) geometrically in
the following manner:  On the unit two-sphere representing the spin directions
corresponding to the projection operators and state above in the
two-dimensional Hilbert space, draw three great circles, one through
$(\theta_0,\phi_0)$ and $(\theta_1,\phi_1)$, one through $(\theta_0,\phi_0)$
and $(\theta_2,\phi_2)$, and one through $(\theta_1,\phi_1)$ and
$(\theta_2,\phi_2)$.  These will generically divide the two-sphere into eight
spherical triangles, four of which are parity reverses of the antipodal four.
Now the conditions (\ref{eq:39})-(\ref{eq:41}) can be represented by geometric
properties of these triangles.

 In particular, the condition given by the real Eq.~(\ref{eq:39}) is equivalent
to the condition that the two great circles through the point
$(\theta_1,\phi_1)$ (which represents $Q$) intersect orthogonally there, so
that each of the eight triangles are right spherical triangles (or the
degenerate limit in which $(\theta_1,\phi_1)$ or the antipodal point to that
coincides with one of the two vertices representing $\rho$ and $R$).  This
condition is satisfied by a three-parameter subset (of measure zero) of the
four-parameter set of directions $(\theta_1,\phi_1)$ and $(\theta_2,\phi_2)$,
assuming that the direction $(\theta_0,\phi_0)$ representing the state is kept
fixed.

 Similarly, the condition given by the complex Eq.~(\ref{eq:40}) is only
satisfied in the degenerate case in which $(\theta_1,\phi_1)$ or its antipode
coincides with either $(\theta_0,\phi_0)$ or $(\theta_2,\phi_2)$; i.e., when
$Q$ or $I-Q$ coincides with either $\rho$ or $R$.  This condition is satisfied
by a discrete family of two-parameter subsets (also of measure zero) of the
four-parameter set of directions $(\theta_1,\phi_1)$ and $(\theta_2,\phi_2)$.

 Finally, the Linearly Positive Histories condition, given by the inequality
(\ref{eq:41}), is equivalent to the condition that none of the eight triangles
have area or solid angle greater than $\pi$ (which is twice the average).
Unlike the other conditions, which give sets of measure zero, this inequality
condition has a positive measure, $(\sqrt{128}-9)/15$, or about $0.154247$, of
the total measure for all possible choices of the two directions
$(\theta_1,\phi_1)$ and $(\theta_2,\phi_2)$ that define the sequence of
projection operators $C(p)$, assuming a measure density for these directions
that is uniform over the two-sphere.

 One point of all these examples is to show that the additional structure of
the conscious world and the corresponding awareness and experience operators in
the quantum world lead to probabilities that need {\it not} be merely
proportional to the ordinary quantum ``probabilities'' in any single set of
possibilities in the same Hilbert space (the ``probabilities'' that I am
claiming are merely fictitious).  Only in the cases of very strong hypotheses,
such as the Commuting Projection Hypothesis, the Orthogonal Projection
Hypothesis, and some of the various Histories Hypotheses, can one get such
proportionalities, since in more general cases one simply has more perceptions
than possibilities in any single set.  Unless one starts from a broader set of
probabilities than those for a single set of possibilities (whether for events
or for histories) in the same Hilbert space, one simply does not get anything
proportional to the true (frequency-type) probabilities for perceptions in
Sensible Quantum Mechanics unless one makes restrictive assumptions about the
additional structure of the expectation values of the awareness or experience
operators.

 One way to circumvent this conclusion is to extend the Hilbert space to a
tensor product of copies of the original Hilbert space and consider ordinary
quantum ``probabilities'' in this larger Hilbert space.  For example, suppose
one has the sequence of $n$ projection operators
$C(h)=\{P(h,1),P(h,2),\cdots,P(h,n)\}$ representing a homogeneous history $h$.
(Note that, unlike in the Sequence of Projections Hypothesis, this $C$ is not
the product of $n$ projection operators in the original Hilbert space, but an
ordered sequence of $n$ projection operators.)  This sequence can now be
regarded as a projection operator on the tensor product of $n$ copies of the
original Hilbert space \cite{I,IL,ILS,S} with the corresponding state
$\sigma^n$ which is the tensor product of $n$ copies of $\sigma$.  Then one can
define a decoherence functional on pairs of atomic histories as
 \begin{eqnarray}
 D(h,h') & = &\sigma^n[C(h')^{\dagger}C(h)]
 \nonumber \\
 & \equiv &\sigma[P(h',1)P(h,1)]\sigma[P(h',2)P(h,2)]
 \cdots\sigma[P(h',n)P(h,n)].
 \label{eq:42}
 \end{eqnarray}

 One can now extend this definition of $C(h)$ for a homogeneous history $h$ to
one for an inhomogeneous history $h$ (a sum of sequences that cannot be written
as a single sequence) by linearity and then define the corresponding
decoherence functional $D(h,h')$ from Eq.~(\ref{eq:42}) by requiring it to be
bilinear in both $C(h)$ and $C(h')$.  This decoherence functional is then not
the standard one that is an expectation value in the state $\sigma$ in the
original Hilbert space, but it is an expectation value in the product state
$\sigma^n$ in the product Hilbert space.  It obviously obeys all the
decoherence conditions of hermiticity, positivity, additivity, and
normalization \cite{GMH,H,I,IL,ILS,S} for any normalized positive state
$\sigma$, thus illustrating how Eq.~(\ref{eq:42}) and its linear extension to
inhomogeneous histories gives a decoherence functional that decoheres for {\it
all} pairs of histories.

 Now, for simplicity, consider the case in which there is only a finite number
$n$ of possible perceptions in the set $M$, each with its corresponding
positive experience operator $E(p)$.  (The case in which there is an infinite
number of possible perceptions would necessitate going to an infinite tensor
product of the original Hilbert space, which will not be done here.)  Assume,
as is the case in which the state is given by a density matrix in a
finite-dimensional Hilbert space, that each $E(p)$ can be written as a sum,
with positive coefficients, of a complete set of orthogonal projection
operators $P(h_p)$:
 \begin{equation}
 E(p)=\sum_{h_p}{\lambda_{h_p} P(h_p)}.
 \label{eq:43}
 \end{equation}
Here we can regard $p$ as an integer between 1 and $n$, inclusive, that labels
the $n$ perceptions in the set $M$.  Now we can regard a basic homogeneous
history $h$ as a particular sequence ${h_1,h_2,\cdots,h_n}$ of the $n$ labels,
giving a particular sequence of projection operators,
$C(h)=\{P(h_1),P(h_2),\cdots,P(h_n)\}$.  Then from the decoherence functional
$D(h,h')$ given by Eq.~(\ref{eq:42}), whose diagonal element $D(h,h)$ can be
considered to be the quantum ``probability''
 \begin{equation}
 P_r(h)\equiv P_r(h_1,h_2,\cdots,h_n)\equiv
D(h,h)=\sigma[P(h_1)]\sigma[P(h_2)]
 \cdots\sigma[P(h_n)],
 \label{eq:44}
 \end{equation}
one {\it can} indeed form linear combinations for the measure
$m(p)=\sigma[E(p)]$ for each perception:
 \begin{equation}
 m(p)=\sum_{h_1,\cdots,h_n}
 {\lambda_{h_p} P_r(h_1,\cdots,h_n)}.
 \label{eq:45}
 \end{equation}

 However, if the number $n$ of perceptions is larger than the square of the
dimension $N$ of the original Hilbert space, then in general one cannot write
the measure for each perception as a linear combination (using coefficients
that depend only on the experience operators and that are independent of the
state itself) of a set of ``probabilities'' that are expectation values for a
single set of possibilities (whose ``probabilities'' add up to one) in a state
in the original Hilbert space.

 Of course, if one adopts a ``many-many-worlds'' interpretation for the quantum
world and assumes the reality of the ``probabilities'' for all sets of
possibilities for some appropriate family defining all of these sets, then this
broader set of ``probabilities'' in suitable cases {\it can} give quantities
that are proportional to the expectation values of the awareness or experience
operators.  Thus one could say that the effect of a quantum state upon the
measures in the conscious world where our experiences lie is proportional to an
appropriate set of ``probabilities'' that can be ascribed to that state in the
quantum world (assuming that the awareness or experience operators are known),
but it can easily be within Sensible Quantum Mechanics that the measures for
the conscious world is not directly given by {\it any single set} of
``probabilities'' (adding up to unity) in the same Hilbert space of the quantum
world.  Unless Sensible Quantum Mechanics takes on a very restricted form, or
unless one extends the Hilbert space sufficiently, the ordinary probabilism
applied to the quantum world is simply inadequate to give directly the measure
for all experiences.

\section{EPR and Schr\"{o}dinger's Cat}

\hspace{.25in}It may be of interest to give a brief analysis of the
Einstein-Podolsky-Rosen (EPR) ``paradox'' \cite{EPR} combined with that of
Schr\"{o}dinger's cat \cite{Sch}.  I shall use a variant of Bohm's modification
\cite{Boh} of the EPR experimental setup to two spin-half atoms in a singlet
state.

 Suppose that the two atoms are moved far apart (while their spins remain
undisturbed in their perfectly anticorrelated singlet state of total angular
momentum zero), and then interactions are made with the atoms' spins in two
spacelike-separated regions of flat Minkowski spacetime, say A and B.  Suppose
that in region A, a perfect nondemolition measurement interaction is made of
the spin of the atom there in the $z$-direction, and suppose that the measuring
device is further coupled to a conscious being so that there results a set of
conscious perceptions $S_{\uparrow}$ corresponding to perceiving that the
experiment has been done, that one is in region A, that the spin direction is
now known by the being, and that the atom was measured to have spin up, and
another set of conscious perceptions $S_{\downarrow}$ corresponding to
perceiving that the experiment has been done, that one is in region A, that the
spin direction is now known by the being, and that the atom was measured to
have spin down.  Here I shall assume an SQMPC theory with the Commuting
Projection Hypothesis (and later the restriction of this to a SQMPPC theory
with the Commuting Product Projection Hypothesis when I discuss a conscious
being in region B with different discrete components to his or her
perceptions), so that the awareness operators $A(S_{\uparrow})$ and
$A(S_{\downarrow})$ for the two sets of perceptions described above are
commuting projection operators which also commute with awareness operators I
shall describe momentarily for certain sets of perceptions of being in region
B.

 I shall further assume for simplicity the idealization that the quantum state
with the experimental setup in region A, including the coupling to the
conscious being, is such that $A(S_{\uparrow})$ and $A(S_{\downarrow})$ have
the same expectation values as commuting projection operators $P_{\uparrow}$
and $P_{\downarrow}$ for the spin of the atom in region A, each multiplied by a
common commuting projection operator $P_{\rm other}$ for other common factors
that lead to the perception that the experiment has been done, that one is in
region A, and that the spin direction is now known by the being.  Thus I am
assuming a perfect correlation between the spin of the atom and whatever (e.g.,
some state in the brain) it is that directly causes the perception that the
spin is up or down.  Furthermore, the idealization I am assuming implies that
whether the spin is up or down has no effect on the expectation value of the
awareness operator (i.e., the measure) for the set of perceptions that the
experiment has been done, that one is in region A, and that the spin direction
is now known by the being.  (For example, I am assuming that no anesthetic is
administered to the being if and only if the atom has spin up, which certainly
could affect the relative measures of the two sets of conscious perceptions by
reducing the measure for $A(S_{\uparrow})$, though in principle SQM is capable
of handling such effects by a more complicated analysis.)

 Now since the pair of atoms was in the singlet state (further assuming that
the perception that the experiment has been done does indeed imply that, though
again SQM could in principle handle the more realistic case in which such
perceptions may be mistaken), the expectation values for $P_{\uparrow}$ and
$P_{\downarrow}$ are identical, and so by our idealization the measures
$\mu(S_{\uparrow})$ and $\mu(S_{\downarrow})$ are also identical, being the
expectation values of  $A(S_{\uparrow})$ and $A(S_{\downarrow})$ that are the
same as those of the commuting projection operators $P_{\uparrow}$ and
$P_{\downarrow}$ multiplied by the common commuting projection operator $P_{\rm
other}$.  Thus in region A there is an equal relative probability to perceive
the atom there having spin up or down, completely independent of what may be
happening in region B.  In other words, under the assumption of local quantum
field theory (which may be only approximately valid if the fundamental theory
is something different, such as string theory), and under the assumption that
the awareness operators for the perceptions that one is in region A are
operators that can be confined to that region (which might be only
approximately valid as well, even under the assumption of local field theory
for the quantum world), there is no superluminal propagation of anything from
region B that can have any effect on anything observed or perceived in the
spacelike-separated region A \cite{P82}.

 Now let us suppose that in region B there is a device to make a nondemolition
measurement interaction with the spin of the atom there in a direction at an
angle $\theta$ from the $z$-direction (which I assume is parallelly propagated
across the flat spacetime from the $z$-direction  in region A).  Suppose that
further there is a ``diabolical device'' \cite{Sch} to poison a cat if and only
if the spin is measured to be down (i.e., if the measuring device, whose record
of the measurement interaction is assumed to have become perfectly correlated
with the atom spin in the $\theta$-direction, has a record indicating that the
spin is down).  For the purposes of the following discussion, divide the cat
into ``head'' and ``body'' (conceptually, not physically; I do not mean to
behead the cat!).  Assume that if the cat is poisoned, that both the head and
the body are dead, but that if the cat is not poisoned, both the head and body
are alive.

 If the projection operator for the atom spin in region B to be up in the
$\theta$-direction is $P_{\rm up}$ and to be down is $P_{\rm down}$, if the
projection operator for the cat head to be alive is $P_{\rm head\ alive}$ and
to be dead is $P_{\rm head\ dead}$, and if the projection operator for the cat
body to be alive is $P_{\rm body\ alive}$ and to be dead is $P_{\rm body\
dead}$ (all of which are assumed to commute), then the density operator for the
atom spin and the cat's property of having its head and body alive or dead
(multiplied by the unit operator for the rest of the total system) is, under
the idealization that the pair of atoms started in the singlet state and that
the coupling to the measuring apparatus and diabolical device were perfect,
proportional to
 \begin{equation}
 P_{\rm up}P_{\rm head\ alive}P_{\rm body\ alive}
 +P_{\rm down}P_{\rm head\ dead}P_{\rm body\ dead}.
 \label{eq:46}
 \end{equation}
Thus there is assumed to be a perfect correlation between the spin of the atom
in region B and the ``liveliness'' of the cat's head and body.

 Now suppose we add a conscious being that perceives that the experiment has
been done as stated (which I shall for simplicity assume excludes the conscious
perceptions of the cat, even if it is still alive; if the cat could perceive
that the experiment has been done and could correctly perceive the spin
direction, presumably it could only perceive that the spin were up, since if
the spin were down, the cat would be dead and presumably would have no
perceptions) and that the being is in region B, and, as for the being perceived
to be in region A, assume the idealization that these perceptions are indeed
perfectly correlated with the state of affairs but are unaffected by whether
the atom spin is up or down.  (For example, the poison is not supposed to kill
this conscious being or otherwise affect the measure of these perceptions of
him or her.)

 Furthermore assume that this conscious being interacts with (e.g., sees) the
liveliness of the cat's head and body.  I now explicitly assume that we have an
SQMPPC theory with the Assumption of Perception Components, one for the
perceived state of the head of the cat and one for the state of the body.

 We now must face the question of what it is about the head and body of the cat
that can lead to different perception components.  The most realistic
idealization seems to be that the SQMPPC theory is such that if the head and
body are alive, they will be perceived to be alive, and if they are dead, they
will be perceived to be dead.  Then different conceivable perceptions could be
in the set $S_{(\rm head\ alive,\ body\ alive)}$ with the components $\{c_{\rm
head\ alive}, c_{\rm body\ alive},\ldots\}$ if both the head and body were
perceived to be alive, with the corresponding awareness operator having an
expectation value equal to that of $P_{\rm head\ alive}P_{\rm body\ alive}$ and
hence being nonzero.  Or, they could be in the set $S_{(\rm head\ dead,\ body\
dead)}$ with components $\{c_{\rm head\ dead}, c_{\rm body\ dead},\ldots\}$ if
both the head and body were perceived to be dead, with the corresponding
awareness operator having an expectation value equal to that of $P_{\rm head\
dead}P_{\rm body\ dead}$ and hence being the same nonzero value, in the
idealization of perfect coupling and with the pair of atoms starting out in the
singlet spin state.  Thus the relative probability of perceiving that the cat
is alive is the same as that of perceiving that the cat is dead, assuming
SQMPPC and the various idealizations above, independent of what may be
happening in region A.

 Other conceivable perceptions of the conscious being in region B that
correctly perceives that the experiment has been done as stated could be in the
set \\$S_{(\rm head\ alive,\ body\ dead)}$ with the components $\{c_{\rm head\
alive}, c_{\rm body\ dead},\ldots\}$ if the head were perceived to be alive but
the body were perceived to be dead, with the corresponding awareness operator
having an expectation value equal to that of $P_{\rm head\ alive}P_{\rm body\
dead}$, or in the set $S_{(\rm head\ dead,\ body\ alive)}$ with components
$\{c_{\rm head\ dead}, c_{\rm body\ alive},\ldots\}$ if the head were perceived
to be dead but the body were perceived to be alive, with the corresponding
awareness operator having an expectation value equal to that of $P_{\rm head\
dead}P_{\rm body\ alive}$.  However, both of these sets of perceptions have
zero expectation value in the idealized state being assumed, so there will be
no perception of a disagreement between the liveliness of the head and body of
the cat with the idealizations being made.  This seems to agree fairly well
with my perception of what has been reported to me, though in my sheltered life
as a theoretical physicist I cannot presently recall any memories of actually
having seen a dead cat myself.

 However, one could imagine an alternative SQMPPC theory to the original one
just described, in which the components of perceptions are not directly coupled
to the eigenstates of the liveliness of the cat's head and body, but, say, to
equal linear combinations of these eigenstates (for either the head or the body
liveliness subsystem as may be the case),
 \begin{eqnarray}
 |{\rm head +}\rangle = & (|{\rm head\ alive}\rangle
 + |{\rm head\ dead}\rangle)/\sqrt{2},\\
 |{\rm head -}\rangle = & (|{\rm head\ alive}\rangle
 - |{\rm head\ dead}\rangle)/\sqrt{2},\\
 |{\rm body +}\rangle = & (|{\rm body\ alive}\rangle
 + |{\rm body\ dead}\rangle)/\sqrt{2},\\
 |{\rm body -}\rangle = & (|{\rm body\ alive}\rangle
 - |{\rm body\ dead}\rangle)/\sqrt{2},
 \end{eqnarray}
or, more precisely, to the corresponding projection operators
 \begin{eqnarray}
 P_{\rm head +} = &
 |{\rm head +}\rangle\langle{\rm head +}|
 \otimes I_{\rm all\ but\ head},\\
 P_{\rm head -} = &
 |{\rm head -}\rangle\langle{\rm head -}|
 \otimes I_{\rm all\ but\ head},\\
 P_{\rm body +} = &
 |{\rm body +}\rangle\langle{\rm body +}|
 \otimes I_{\rm all\ but\ body},\\
 P_{\rm body -} = &
 |{\rm body -}\rangle\langle{\rm body -}|
 \otimes I_{\rm all\ but\ body},
 \end{eqnarray}
where $I_{\rm all\ but\ head}$ is the identity operator for all of the system
except for the head liveliness subsystem with its states $|{\rm head +}\rangle$
and $|{\rm head -}\rangle$, and similarly $I_{\rm all\ but\ body}$ is the
identity operator for all of the system except for the body liveliness
subsystem.

 Then different conceivable perceptions could be in the set $S_{(\rm head +,
body +)}$ with the components $\{c_{\rm head +}, c_{\rm body +},\ldots\}$ if
both the head and body were perceived to be in their + states, with the
corresponding awareness operator having an expectation value equal to that of
$P_{\rm head +}P_{\rm body +}$; in the set $S_{(\rm head -, body -)}$ with
components $\{c_{\rm head -}, c_{\rm body -},\ldots\}$ if both the head and
body were perceived to be in their $-$ states, with the corresponding awareness
operator having an expectation value equal to that of $P_{\rm head -}P_{\rm
body -}$; in the set $S_{(\rm head +, body -)}$ with the components $\{c_{\rm
head +}, c_{\rm body -},\ldots\}$ if the head were perceived to be in the +
state but the body were perceived to be in the $-$ state, with the corresponding
awareness operator having an expectation value equal to that of $P_{\rm head
+}P_{\rm body -}$; or in the set $S_{(\rm head -, body +)}$ with components
$\{c_{\rm head -}, c_{\rm body +},\ldots\}$ if the head were perceived to be in
its $-$ state but the body were perceived to be in its + state, with the
corresponding awareness operator having an expectation value equal to that of
$P_{\rm head -}P_{\rm body +}$.  All of these awareness operators would have
the same nonzero expectation value in the idealizations of perfect coupling and
of the pair of atoms starting out in the singlet spin state.

 Thus in this alternative SQMPPC theory and in the idealized experiment being
described, there would be no correlation between the perception of the states
of the cat's head and body, when one averages over all sets of perceptions,
weighted by their measures, in which the experiment is perceived to have
occurred (assuming that this component of the perception is indeed perfectly
correlated with whether or not the experiment occurred as stated).  Presumably
we might describe the conscious being as being confused if he or she perceives
a disagreement between the states of the cat's head and body.  In the original
SQMPPC theory described above, the measure was zero (under the idealized
assumptions) for such ``confused'' perceptions, so all perceptions with
positive measure were ``unconfused,'' but in the alternative SQMPPC the
weighted fraction of unconfused perceptions was only one half (and would have
been only $2^{1-n}$ if one had conceptually divided the cat into $n$ parts that
were all assigned + and $-$ states to which the conscious perceptions were
perfectly correlated).

 The comparison of these two conceivable SQMPPC theories shows that if it is
desired that perceptions $p$ be all unconfused in idealized cases, their
components $c_i(p)$ should have their corresponding projection operators having
expectation values equal to those of $P[c_i(p)]$ coupled to things (e.g.,
states of parts of the cat) that are entirely correlated (e.g., the
``liveliness'' property of being alive or dead, rather than the + or $-$
properties) in these cases.  These preferred projection operators for
unconfused conscious perceptions are similar to the Information Basis of States
for quantum measurements \cite{P86}, except that no claim is made in SQM that
the operators associated with perceptions form a complete basis.

 However, it still is somewhat confusing to me why in idealized cases our
perceptions actually seem to be rather unconfused,
why the original rather than the alternative SQMPPC theory seems more accurate
(or more likely to make our unconfused perceptions typical).  One might argue that if
they were not unconfused, then we could not act coherently and so would not
survive.  This would seem to be a good argument only if our perceptions really
do affect our actions in the quantum world and are not just epiphenomena that
are determined by the quantum world without having any effect back on it.
Another argument similarly suggesting that explanations might be simpler if
conscious perceptions acted back on the quantum world will be mentioned briefly
in the Conclusions.  But on the other hand, it is not obvious how perceptions
could affect the quantum world in a relatively simple way in detail (though it
is easy to speculate on general ways in which there might be some effect; see
\cite{Page} and the Conclusions below).  So although it appears to be
unexplained, it conceivably could be that conscious perceptions do not affect
the quantum world but are determined by it in just such a way that in most
cases they are not too confused.
To mimic Einstein, I might say, ``The most confusing thing about perceptions
is that they are generally unconfused.''

 As an aside, I should say that although epiphenomenalism
seems to leave it mysterious {\it why} typical perceptions are unconfused,
I do not think it leaves it mysterious {\it that} perceptions occur,
despite a na\"{\i}ve expectation that the latter is also mysterious.
The na\"{\i}ve argument is that if the conscious world has no effect
on the quantum world (usually called the physical world \cite{Pen,Chal},
in contrast to my use of that term to include both the quantum world
and the conscious world), and if the development of life in the
quantum world occurs by natural selection,
the development of consciousness would have no effect
on this natural selection and so could not be explained by it.

 Nevertheless, one can give an answer analogous to that I have heard was given by
the former Fermilab Director Robert Wilson when he was was asked
by a Congressional committee what Fermilab contributed to the defense of the nation:
"Nothing.  But it helps make the nation worth defending."
Similarly, if epiphenomenalism is correct, consciousness may contribute nothing
to the survival of the species, but it may help make certain species worth surviving.
More accurately, it may not contribute to the evolution of complexity,
but it may select us (probably not uniquely) as complex organisms
which have typical perceptions.  Then our consciousness would not be surprising,
because we are selected simply as typical conscious beings.

 This selection as typical conscious beings might also help explain
why we can do highly abstract theoretical mathematics and physics
that does not seem to help us much with our survival as a species.
If we are selected by the measure of our consciousness,
and if that is positively correlated with a certain kind of complexity
that is itself correlated with the ability to do theoretical mathematics and physics,
then it would not be surprising that we can do this better than the average hominid
that survives as well as we do (say averaging over all the Everett many worlds).

 Returning to a consideration of the EPR experiment,
the perceptions by conscious beings in the spacelike-separated regions A and B
occur independently in SQM (under the idealizations of local quantum field
theory and of the assumption that the relevant awareness operators are confined
to either region A or B) and do not show any of the EPR correlations.  To
perceive the EPR correlations between regions A and B, one needs perceptions
with awareness operators whose expectation values are affected by what is
happening in both regions.  One obvious way is to have these awareness
operators in a region C which is to the causal future of both regions A and B.
For example, a signal could be sent from each region, a signal that is perfectly
correlated with the result of the spin measurement in that region
(another idealization made for simplicity). 
Then one can imagine conscious perceptions of
a being in C (meaning that the corresponding awareness operators can be
localized to that region) with one component of each determined by the signal
from A and the other component determined by the signal from B.

 For example, there could be the set $S_{\uparrow\uparrow}$ of perceptions in
which both signals indicate that the corresponding spin is up in the measured
directions (which differed by the angle $\theta$), the set
$S_{\uparrow\downarrow}$ of perceptions that the spin in A is up and that the
spin in B is down, the set $S_{\downarrow\uparrow}$ of perceptions that the
spin in A is down and that the spin in B is up, and the set
$S_{\downarrow\downarrow}$ of perceptions that the spin in both A and B is
down.  The awareness operators for these four sets of perceptions could have
expectation values, ideally, the same as those of $P_{\uparrow}P_{\rm up}$,
$P_{\uparrow}P_{\rm down}$, $P_{\downarrow}P_{\rm up}$, and
$P_{\downarrow}P_{\rm down}$, respectively, each multiplied by a common
commuting projection operator $P_{\rm other C}$ for other common factors that
lead to the perception that the experiment has been done, that one is in region
C, and that both spin directions are now known as a result of receiving signals
from both regions A and B.

 Now one can readily calculate that if the pair of atoms starts in the singlet
spin state, then under the idealizations above, the sets of perceptions
$S_{\uparrow\uparrow}$ and $S_{\downarrow\downarrow}$ each have the same
measure, which is $\tan^2{\theta/2}$ of the measure for each of the sets
$S_{\uparrow\downarrow}$ and $S_{\downarrow\uparrow}$.  Thus if $\theta=0$,
there will be no measure for the first two sets of perceptions, and the
conscious being in C will necessarily perceive that the two spins are opposite,
the perfect EPR anticorrelation, even though the measure for each of the last
two sets of perceptions is equal so that one cannot uniquely predict whether it
will be the spin in A or the spin in B that is perceived to be up in all
perceptions with nonzero measure.

 However, since one sees that the EPR correlations between regions A and B can
be perceived by conscious beings (assuming that their awareness operators can
be fairly well localized to where we can thus define these beings to be) only
if they are to the causal future of both A and B (e.g., in region C), we do not
have any superluminal propagation of anything in SQM if it is based on local
quantum field theory.  (Of course, the ultimate replacement of quantum field
theory, e.g., by string theory, may eliminate this locality property of the
quantum world, except as some sort of approximation in suitable circumstances.)
 Furthermore, just as one may conclude concerning quantum mechanics in the
Everett interpretation \cite{P82}, so too the Einstein-Podolsky-Rosen physical
reality is completely described by Sensible Quantum Mechanics, contrary to the
claim that Einstein, Podolsky, and Rosen \cite{EPR} made about quantum
mechanics when all they had was the Copenhagen interpretation.

\section{Questions and Speculations}

\hspace{.25in}One can use the framework of Sensible Quantum Mechanics to ask
questions and make speculations that would be difficult without such a
framework.  I shall here give some examples, without intending to imply that
Sensible Quantum Mechanics itself, even if true, would guarantee that these
questions and speculations make sense, but it does seem to allow circumstances
in which they might.

 First, in the model of quantum field theory on a classical spacetime with no
symmetries, and with a quantum state having well-localized human brains on some
Cauchy hypersurface labeled by time $t$, one might ask whether it is possible
to have two quite different perceptions, say $p$ and $p'$, in nearly the same
Everett world in the sense of having the $f(p,p')$ of Eq.~(\ref{eq:P5}) near
unity, and giving $E(p)$ and $E(p')$ both with the same preferred time $t_p=t$
and both localized (by the rather {\it ad hoc} prescription of Section 8) in
balls in the same brain.  In other words, can one brain have two different
(maximal) perceptions in the same world at the same time, each not aware of the
other?  Unless we are solipsists (or unless we adopt the Orthogonal Projection
Hypothesis, in which case we say that different perceptions all occur in
different Everett worlds), we generally believe this is possible for two
separate brains, but would one brain be sufficient?  Furthermore, if it is
possible, can the two balls (corresponding to $p$ and $p'$ respectively) be
overlapping spatially, or need they be separate regions in the brain?

 Second, one might ask whether and how the sum (or integral) of the measures
(or measure densities) $m(p)$ associated with an individual brain region at the
time $t$ depends on the brain characteristics.  One might speculate that it
might be greater for brains that are in some sense more intelligent, so that in
a crude sense brighter brains have more perceptions.  This could explain why
you do not perceive yourself to be an insect, for example, even though there
are far more insects than humans.

 Third, one might conjecture that an appropriate measure on perceptions might
give a possible explanation of why most of us perceive ourselves to be living
on the same planet on which our species developed.  This observation might seem
surprising when one considers that we may be technologically near the point at
which we could leave Earth and colonize large regions of the Galaxy \cite{Dys},
presumably greatly increasing the number of humans
beyond the roughly $10^{11}$ that are believed to have lived on Earth.
If so, why don't we have the perceptions of
one of the vast numbers of humans that may be born away from Earth?
One answer is that some sort of doom is likely
to prevent this vast colonization of the Galaxy from happening
\cite{C,Le,N,G},
%[46-49],
though these arguments are not conclusive \cite{KKP}.  Although I would not be
surprised if such a doom were likely, I would na\"{\i}vely expect it to be not
so overwhelmingly probable that the probability of vast colonization would be
so small as is the presumably very small ratio of the total number of humans
who could ever live on Earth to those who could live throughout the Galaxy if
the colonization occurs.  Then, even though the colonization may be unlikely,
it may still produce a higher measure for conscious perceptions of humans
living off Earth than on it.

 However, another possibility is that colonization of the Galaxy is not too
improbable, but that it is mostly done by self-replicating computers or
machines who do not tolerate many humans going along, so that the number of
actual human colonizers is not nearly so large as the total number who {\it
could} live throughout the Galaxy if the computers or machines did not dominate
the colonization.  If the number of these computers or machines dominate humans
as ``intelligent'' beings (in the sense of having certain information-processing
capabilities), one might still have the question of why we perceive ourselves
as being humans rather than as being one of the vastly greater numbers of such
machines.  But the explanation might simply be that the {\it weight} of
conscious perceptions (the sum or integral of the $m(p)$'s corresponding to the
type of perceptions under consideration) is dominated by human perceptions,
even if the {\it number} of ``intelligent'' beings is not.  In other words,
human brains may be much more efficient in producing conscious perceptions than
the kinds of self-replicating computers or machines which may be likely to
dominate the colonization of the Galaxy.  If such machines are more
``intelligent'' than humans in terms of information-processing capabilities and
yet are less efficient in producing conscious perceptions, our perceptions of
being human would suggest that the measure of perceptions is not merely
correlated with ``intelligence.''  (On the other hand, if the measure of
perceptions is indeed strongly correlated with ``intelligence'' in the sense of
information-processing capabilities, perhaps it might be the case that Galactic
colonization is most efficiently done by self-replicating computers or machines
that are not so ``intelligent'' as humans.  After all, insects and even
bacteria have been more efficient in colonizing a larger fraction of Earth than
have humans.)

 It might be tempting to take the observations that these speculations might
explain (our perceptions of ourselves as human rather than as insect, and our
perceptions of ourselves as humans on our home planet) as evidence tending to
support the speculations.  One could summarize such reasoning as a
generalization of the Weak Anthropic Principle
\cite{D,Ca,CR,Ro,Da,BT,Les}
%[56-62]
that might be called the {\it Conditional Aesthemic Principle} (CAP):  given
that we are conscious beings, our conscious perceptions are likely to be
typical perceptions in the conscious world with its measure.

 If one uses the dual typicality $T_d(p)$ defined by Eq.~(\ref{eq:14d}) as an
indication of how ``likely'' a perception is, one can say that there is a 99\%
likelihood that $T_d(p)\geq 0.01$.  For example, if one restricts oneself to
the perception of a continuous variable for which the measure density has a
gaussian distribution, then at the 99\% likelihood level, the variable should
be between about 0.0062666117 and about 2.8070337863 standard deviations from
the mean.  Values closer to the mean are ``too good to be likely,'' and values
further from the mean are ``too bad to be likely,'' at least at the 99\%
likelihood level.

 In addition to the typicality $T(p)$ defined by Eq.~(\ref{eq:13}) and the dual
typicality $T_d(p)$ defined by Eq.~(\ref{eq:14d}) as indications of how
``likely'' a perception is, the discussion above suggests the usefulness of
other measures of the typicality or likelihood of a perception.  Unfortunately,
there is the apparent arbitrariness of their definition.  If one makes a rather
{\it ad hoc} definition of typicality and then finds that one's perception is
atypical with respect to this definition, one may have grounds to be sceptical
that it really is evidence against the detailed Sensible Quantum Mechanics
theory that predicts the low typicality-thus-defined of one's perception.  For
example, I have the perception of having been one of the last $10^{-7}$ or so
of those born in 1948 (since I was born about 90 minutes before midnight on
December 31 according to the local time, ten time zones west of Greenwich, in
the lowly-populated Territory of Alaska, fairly near the point $\cal B$
of Figure 8.3 on page 212 of my Ph.D. advisor's textbook \cite{MTW},
with only Hawaii providing a significant population in that
or any other time zone further west),
so I am atypical in that regard, but it is probably not so surprising that after living
over 40 years I have finally found some particular detail about myself that by
itself might appear unusual.

 I also noticed recently that the fourth (and, I would guess, last) in the
sequence of at least four Mersenne primes given by the (non-Mersenne) prime
seed $N_0=2$ and the recursion relation $N_{n+1}=2^{N_n}-1$ is within one-half
of one percent of the inverse gravitational fine structure constant for the
proton (the square of the ratio of the Planck mass to the proton mass).  (The
logarithms of these two large numbers agree to about one part in $19000\pm
500$.)  If the inverse of the gravitational fine structure constant is an
environmental constant that varies from component to component of the quantum
state of the universe, then I would expect it to be rather atypical to perceive
it to be so close to $2^{127}-1$, but since there are presumably so many other
unusual things that I could have found instead (and which indeed Eddington
\cite{Ed} did find and claim to have found a theory for), I would think that
surely it is just another numerical coincidence and thus that the apparent
resulting ``atypicality'' is just an artifact of an {\it ad hoc} definition for it.

 I could also cite various other perceptions that na\"{\i}vely might seem atypical,
such as my frequent remembrance that in 1992
the U.S., Canada, and the new Russian Republic
had ages in years that were all perfect cubes (216, 125, and 1 respectively),
my noticing that an integer often considered unlucky in Western culture
is the only positive integer fourth root of the sum of two successive
positive square integers,
and my fascination with the fact that there is a mystery word
(for which I have a long standing offer of \$10 for the first person who can find it)
whose modern definition in the British {\em Oxford English Dictionary}
gives a certain quantity, but whose definition
in the American {\em Webster's Third International Dictionary}
gives a different quantity, larger by the ratio of about \\
1.0006551997916815342586087238773730720409883464864441784178751363.

 Part of the explanation for such `unusual' perceptions
is what I have called the Attention Effect (see the last paper in \cite{Page}).
This is the fact that unusual events attract our attention, so that we tend
to focus on them and spend a longer time being conscious of them.
The measure for a set of perceptions presumably increases with the time
(when that approximate concept is applicable) spent having them,
so events that attract our attention for longer periods of time
will presumably lead to sets of conscious perceptions having greater measure
and hence being less atypical than we might have thought..

 Therefore, one should be very careful in using the Conditional Aesthemic
Principle, even though it might be useful in explaining certain features of our
perceptions that might more na\"{\i}vely be thought to be surprising.

\section{Conclusions}

\hspace{.25in}In conclusion, I am proposing that Sensible Quantum Mechanics is
the best framework we have at the present level for understanding conscious
perceptions and the interpretation of quantum mechanics.  Of course, the
framework would only become a complete theory once one had the set $M$ of all
perceptions $p$, the awareness operators $A(S)$, and the quantum state $\sigma$
of the universe (and preferably also the prior measure $\mu_0(S)$ in order to
test the theory and compare it with others).

 Even such a complete theory of the quantum world and the conscious world
affected by it need not be the ultimate simplest complete theory of the
combined physical world.  There might be a simpler set of unifying principles
from which one could in principle deduce the perceptions, awareness operators,
and quantum state, or perhaps some simpler entities that replaced them.  For
example, although in the present framework of Sensible Quantum Mechanics, the
quantum world (i.e., its state), along with the awareness operators, determines
the measure for perceptions in the conscious world, there might be a reverse
effect of the conscious world affecting the quantum world to give a simpler
explanation than we have at present of the coherence of our perceptions (as
pondered in Section 11) and of the correlation between will and action (why my
desire to do something I feel am capable of doing is correlated with my
perception of actually doing it, i.e., why I ``do as I please'').  If the
quantum state is partially determined by an action functional, can desires in
the conscious world affect that functional (say in a coordinate-invariant way
that therefore does not violate energy-momentum conservation)?  Such
considerations may call for a more unified framework than Sensible Quantum
Mechanics, which one might call Sensational Quantum Mechanics \cite{Page}.
 Such a more unified framework need not violate the limited assumptions of
Sensible Quantum Mechanics, though it might do that as well and perhaps reduce
to Sensible Quantum Mechanics only in a certain approximate sense.

 To explain these frameworks in terms of an analogy, consider a classical model
of spinless massive point charged particles and an electromagnetic field in
Minkowski spacetime.  Let the charged particles be analogous to the quantum
world (or the quantum state part of it), and the electromagnetic field be
analogous to the conscious world (the set of perceptions with its measure
$\mu(S)$).  At the level of a simplistic materialist mind-body philosophy, one
might merely say that the electromagnetic field is part of, or perhaps a
property of, the material particles.  At the level of Sensible Quantum
Mechanics, the charged particle worldlines are the analogue of the quantum
state, the retarded electromagnetic field propagator (Coulomb's law in the
nonrelativistic approximation) is the analogue of the awareness operators, and
the electromagnetic field determined by the worldlines of the charged particles
and by the retarded propagator is the analogue of the conscious world.  (Here one can see that
this analogue of Sensible Quantum Mechanics is valid only if there is no free
incoming electromagnetic radiation.)  At the level of Sensational Quantum
Mechanics, at which the conscious world may affect the quantum world, the
charged particle worldlines are partially determined by the electromagnetic
field through the change in the action it causes.  (This more unified framework
better explains the previous level but does not violate its description, which
simply had the particle worldlines given.)  At a yet higher level, there is the
possibility of incoming free electromagnetic waves, which would violate the
previous frameworks that assumed the electromagnetic field was uniquely
determined by the charged particle worldlines.  (An analogous suggestion for
intrinsic degrees of freedom for consciousness has been made by Linde
\cite{Lin90}.)  Finally, at a still higher level, there might be an even more
unifying framework in which both charged particles and the electromagnetic
field are seen as modes of a single entity (e.g., to take a popular current
speculation, a superstring).

 Therefore, although it is doubtful that Sensible Quantum Mechanics is the
correct framework for the final unifying theory (if one does indeed exist), it
seems to me to be a move in that direction that is consistent with what we
presently know about consciousness and the physical world.  At least it seems
to be an augmentation of ordinary quantum mechanics (without the collapse
postulate) that cannot be criticized as being incomplete for not predicting
(when the framework is fleshed out into a complete SQM theory) precisely what
happens during observations.

\section*{Acknowledgments}

\hspace{.25in}I consciously perceive a gratitude for having had fruitful
discussions or correspondence on this general subject with
David Albert, Warren Anderson, Andrei Barvinsky, Dick Bond, Arvind Borde,
David Boulware, Patrick Brady, Howard Brandt, Jeremy Butterfield, 
Bruce Campbell, Brandon Carter, David Chalmers, Sidney Coleman,
James Cushing, David Deutsch, Daniel Dennett, Bryce DeWitt,
Fay Dowker, Detlef D\"{u}rr, Bernard d'Espagnat,
Willy Fischler, Gordon Fleming, Finola Fogarty,
Valeri Frolov, Murray Gell-Mann, Farhad Ghoddoussi,
Shelly Goldstein, Richard Gott, Robert Griffiths,
Jonathan Halliwell, Jim Hartle, Stephen Hawking,
Geoff Hayward, Gary Horowitz, Bei-Lok Hu, Viqar Husain, Werner Israel,
Nick Kaiser, Renata Kallosh, Adrian Kent, Christof Koch,
Rocky Kolb, Tom\'{a}\v{s} Kopf, Pavel Krtou\v{s}, Karel Kucha\v{r},
John Leslie, Andrei Linde, Michael Lockwood, Robb Mann, Donald Marolf,
Richard Matzner, David Mermin, Artur Mezhlumian, Ivan Muzinich,
Roland Omn\`{e}s, Roger Penrose, Philip Pearle, 
John Polkinghorne, John Preskill, Richard Price, Raghu Rangarajan,
David Salopek, Simon Saunders, John Schwartz, Larry Shepley,
Abner Shimony, Euan Squires (now deceased),
Henry Stapp, Andy Strominger, Lenny Susskind, Kip Thorne, Bill Unruh,
Lev Vaidman, Alex Vilenkin, Bob Wald, Steven Weinberg, John Wheeler,
and various others whose names aren't in my immediate perception.
(Not all of these appear to be in any single perception, at least to any great degree,
since I apparently cannot be consciously aware of so many people at once,
but at different `times' I have presumably been conscious of the infuence of
each of the corresponding persons.)
Of course, none of them are to blame if Sensible Quantum Mechanics is actually
senseless.  I am especially grateful to my wife Cathy for helping me become
more aware of my feelings.  The numerical calculations were performed with the
aid of Mathematica.  Finally, financial support has been provided by the
Natural Sciences and Engineering Research Council of Canada.

\end{document}